\documentclass[a4paper,11pt]{article}
\pdfoutput=1
\usepackage{jheppub} 
\usepackage[T1]{fontenc}
\usepackage[T1]{fontenc}
\usepackage[table]{xcolor}
\usepackage[caption=false]{subfig}
\usepackage{epsfig}
\usepackage{colortbl}
\usepackage[T1]{fontenc}
\usepackage{tikz-feynman}
\usepackage{hepnames}
\usepackage{adjustbox}
\usepackage{textcomp}
\usepackage{slashed}
\newcommand{\bea}{\begin{eqnarray}}
\newcommand{\eea}{\end{eqnarray}}
\usepackage{xcolor}
\newcommand\ccg[1]{\cellcolor{blue!5!red!5!green!5}{#1}} 
\newcommand\ccw[1]{\cellcolor{blue!5}{#1}}
\newcommand*\circi[1]{\tikz[baseline=(char.base)]{\node[shape=circle,fill=none,draw=black,path fading=none,inner sep=1.7pt] (char) {\footnotesize #1};}}
\newcommand*\circii[1]{\tikz[baseline=(char.base)]{\node[shape=circle,fill=none,draw=black,path fading=none,inner sep=1.0pt] (char) {\footnotesize #1};}}
\newcommand*\circiii[1]{\tikz[baseline=(char.base)]{\node[shape=circle,fill=none,draw=black,path fading=none,inner sep=0pt] (char) {\footnotesize #1};}}
\newcommand*\circiv[1]{\tikz[baseline=(char.base)]{\node[shape=circle,fill=none,draw=black,path fading=none,inner sep=0.3pt] (char) {\footnotesize #1};}}
\usepackage[font=small]{caption}
\usepackage[font={small},labelfont={bf}]{caption}
\usepackage{url}
\definecolor{MyDarkBlue}{rgb}{0.1, 0.1, 0.8}
\definecolor{SBlue}{rgb}{0.2, 0.4, 0.7} 
\definecolor{MyLightBlue}{rgb}{0.22,0.51,0.9}
\definecolor{MyGreen}{rgb}{0.0, 0.5, 0.0}
\definecolor{BrickRed}{rgb}{0.8, 0.25, 0.33}
\usepackage{hyperref}
\hypersetup{colorlinks, citecolor=BrickRed,linkcolor=MyDarkBlue, urlcolor=MyGreen}
\title{\boldmath Pseudo-FIMP dark matter in presence of a SIMP}
\author[a]{Subhaditya Bhattacharya, }
\author[a]{Dipankar Pradhan, }
\author[a,b]{Jahaan Thakkar.}
\affiliation[a]{Department of Physics, Indian Institute of Technology Guwahati,\\North Guwahati, Assam-781039, India,}
\affiliation[b]{Department of Theoretical Physics, Tata Institute of Fundamental Research,\\Mumbai, Maharashtra-400005, India,}
\emailAdd{subhab@iitg.ac.in}
\emailAdd{d.pradhan@iitg.ac.in}
\emailAdd{jahaan.thakkar@tifr.res.in}
\abstract{Pseudo-feebly Interacting Massive Particle (pFIMP) has been postulated in two component dark matter (DM) scenarios, where it has feeble interaction with 
the visible sector, but sizeable one with a thermal bath partner. In this work, we study the possibility and dynamics of pFIMP in presence of a Strongly Interacting 
Massive Particle (SIMP), which is well known to solve too-big-to-fail and core-vs-cusp problems. Our analysis is primarily model-independent via solving coupled 
Boltzmann equations, with negligible DM-DM conversion adhering to pure SIMP-FIMP limit, and then with larger DM-DM conversion rate pertaining to SIMP-pFIMP limit. 
We also illustrate the simplest model yielding pFIMP-SIMP set-up having two scalars stabilised under $\mathbb{Z}_2\otimes \mathbb{Z}_3$ symmetry, 
and explore the accessible parameter space after addressing relic density, unitarity, self interaction constraints etc. pFIMP detectability is limited in such circumstances, 
but possible via a thermal DM loop when the SIMP has a visible sector interaction via light mediator.}
\keywords{Models for Dark Matter, Particle Nature of Dark Matter, Specific BSM Phenomenology, Dark Matter at Colliders.}
\makeatletter
\gdef\@fpheader{}
\makeatother
\begin{document}
\maketitle
\flushbottom
\section{Introduction}
Our universe is full of mysteries. The presence of a non-luminous matter component, known as Dark Matter (DM) \cite{Zwicky:1937zza, Zwicky:1933gu} is one of them. 
Different kinds of astrophysical and cosmological observations, like Galaxy rotation curves \cite{Sofue:2000jx}, Bullet Cluster \cite{Hayashi:2006kw}, CMBR anisotropic spectrum \cite{Komatsu:2014ioa} suggest that DM is present and possesses 26.8\% of the energy budget of the Universe and this gives rise to one of the most important observables 
related to DM, called the relic density, expressed as $\rm\Omega_{DM} h^2=0.1200\pm 0.0012$ \cite{Planck:2018vyg}, where $\Omega=\rho/\rho_c$ is cosmological density 
and $h$ is reduced Hubble constant. 

What constitutes the DM is a key question, which has involved a lot of researches and possibilities so far. Particle DM has been very popular since it can judiciously 
explain almost all the observations related to DM. However, none of the observations can pinpoint what kind of particle it is, although some broad characteristics like 
electromagnetic charge neutrality, massiveness and stability at the scale of Universe's life time are assigned to them. Further debate erupts on the question whether the 
DM is composed of a single type of particle or have different particles of different nature, adhering to the 
broad characteristics of DM. Again, the possibilities here are infinite and the only relevant constraint comes from the Bullet cluster observation, which restricts the DM-DM 
self-interaction \cite{Clowe:2006eq}. One thing we know for sure, that Standard Model (SM) of particle physics do not contain any DM and requires extension to accommodate 
one (or more) DM, protected by a symmetry different from that of SM gauge symmetries.

DM characteristics mainly depend on its production mechanism. There are two possibilities (broadly) depending on whether 
it was in thermal bath during reheating or not. Those which were in 
equilibrium with bath and freezes out at some point with expanding universe are called thermal DM, and those which were not in equilibrium and produced via out-of equilibrium processes to freeze in, are called non-thermal ones. The most popular thermal DM candidates are weakly interacting massive particle (WIMP)\cite{Roszkowski:2017nbc,Kamionkowski:1990ni, Gondolo:1990dk,Jungman:1995df,Edsjo:1997bg,Bottino:1993zx} and strongly interacting massive particle (SIMP) \cite{Hochberg:2014dra, Hochberg:2014kqa, Hochberg:2015vrg, Tulin:2017ara, Choi:2018iit, Bhattacharya:2019mmy, Barman:2021ugy, Kamada:2022zwb}, while a well studied non-thermal possibility gives rise to feebly interacting massive particle (FIMP) \cite{Hall:2009bx,Yaguna:2011qn, Bernal:2017kxu, DEramo:2020gpr,Chakrabarty:2022bcn, Biswas:2016iyh} type of DM. Multipartite DM can be constituted by any 
combination of them. The most studied frameworks involve similar DM components, like WIMP-WIMP \cite{Bhattacharya:2013hva, Bhattacharya:2016ysw}, 
FIMP-FIMP \cite{Pandey:2017quk, PeymanZakeri:2018zaa}, and SIMP-SIMP \cite{Ho:2022erb, Choi:2021yps}. WIMP-FIMP \cite{DuttaBanik:2016jzv, Bhattacharya:2021rwh}
combination is possible when the interaction between DM components is also feeble. However, if we have sizeable 
interaction between the DM components, then it gives rise to an interesting possibility of a pseudo FIMP (pFIMP) \cite{Bhattacharya:2022dco, Bhattacharya:2022vxm,Bhattacharya:2024nla} type DM. 
How ? Then FIMP, in spite of having feeble interaction with the visible sector, reaches thermal equilibrium thanks to the large conversion with the WIMP partner, 
and is called pseudo-FIMP (pFIMP). pFIMP then undergoes freeze-out similar to WIMP. However the freeze-out characteristics depend on the WIMP 
partner, making it a special/different DM component.  

We wish to take the study of pFIMP one step ahead. We show here how pFIMP can be realised in presence of a SIMP like DM component. 
The freeze out characteristics of pFIMP then depends very much on those of its SIMP partner. We establish this in a model independent analysis 
via solving coupled Boltzmann Equations, so that it is applicable to any kind 
of pFIMP-SIMP combination, immaterial to their particle properties. Second, we establish the same via a concrete model example involving 
two scalar fields, one real, and the other complex, stabilized by $\mathbb{Z}_2\otimes \mathbb{Z}_3$ symmetry. 
This is arguably the simplest scenario yielding pFIMP-SIMP model. The search strategies for DM known so far, like direct detection (DD), indirect detection (ID), or collider searches are difficult to reveal such DM combinations unless the SIMP is accessible to the detector via light mediator.

This article is organized as follows: sec\,.~\ref{sec:2} provides a brief discussion on a single component SIMP, sec\,.~\ref{sec:3} explores the dynamics and phenomenology of pFIMP in a model-independent manner, and sec\,.~\ref{mod:-simp-pfimp} presents a model example. Finally, we summarize in sec\,.~\ref{sec:5}. Appendices~\ref{app:A}, \ref{app:B}, \ref{app:C}, and \ref{app:D} provide additional details that were omitted in the main text.
\section{A brief discussion on the simplest SIMP model}
\label{sec:2}
For WIMP, the DM freeze out depends on the thermal average of annihilation cross-section ($2_{\rm DM} \to 2_{\rm SM}$). For SIMP, 
DM freeze out (FO) occur mainly via self-interaction like $3_{\text{DM}}\to2_{\text{DM}}$ \cite{Bhattacharya:2019mmy} 
(or $4_{\text{DM}}\to2_{\text{DM}}$ \cite{Bernal:2015xba}) annihilations within the dark sector, while $2_{\rm DM} \to 2_{\rm SM}$ annihilation is suppressed. 
The equilibration of SIMP to thermal bath is maintained by scattering with SM particles (DM+SM $\to$ DM+SM), the rate of which is given by 
$\Gamma=n_{eq}\langle \sigma v \rangle$, which can be larger than the Hubble expansion rate ($H$) due to the large relativistic SM number density ($n_{eq}$), 
in spite of the small annihilation cross-section to SM.  The hierarchy of interactions that keeps SIMP in thermal bath without heating up the dark sector is therefore, 
\bea
\Gamma_{\rm DM+SM\to DM+SM} \gtrsim  \Gamma_{3_{\rm DM}\to 2_{\rm DM}}  \gtrsim  \Gamma_{2_{\rm DM}\to 2_{\rm SM}}\,.
\eea
However for the five body ($3_{\text{DM}}\to2_{\text{DM}}$) or six body depletion process ($4_{\text{DM}}\to2_{\text{DM}}$) that governs SIMP freeze out, 
the DM mass needs to be adjusted to account for the larger phase space volume and is of the order of MeV to address correct relic density. 
The coupling for such a number-changing process within the dark sector must also be significantly large to take into account of the phase space suppression. 
Hence, such a class of DM is called Strongly Interacting Massive Particle (SIMP) 
\cite{Lee:2015gsa, Bernal:2015xba, Hochberg:2015vrg, Choi:2015bya, Bhattacharya:2019mmy}. We may stress here again that 
strong interaction here doesn't refer to strong interaction between dark and SM particles, but it is within the dark sector particles. 

Concerning SIMP, the simplest possibility emerges when we take a complex scalar singlet $(\chi)$, which transforms under $\mathbb{Z}_3$ 
symmetry. Corresponding Lagrangian is \cite{Bhattacharya:2019mmy},
\bea
\mathcal{L}=\mathcal{L}_{\text{SM}}+|\partial_{\mu}\chi|^2-m_{\chi}^2|\chi|^2-\lambda_{\chi}|\chi|^4-\lambda_{\chi H}\left(H^{\dagger}H-\dfrac{v^2}{2}\right)|\chi|^2-\dfrac{\mu_3}{2}\left(\chi^3+{\chi^{\star}}^3\right)\,,
\label{eq:lag}
\eea
where $\mu_3, \lambda_\chi, \lambda_{\chi H}>0$. Note here that the above Lagrangian is valid for any charge of $\chi$ or $\chi^*$ as $\omega,\omega^2$, where $\omega^3=1$.
\paragraph{$\bullet$ Unitarity and Perturbativity\\}
Using the unitarity of the S-matrix, {\it i.e\,.~}the optical theorem, the maximum value of the inelastic cross-section for an identical particle is \cite{Griest:1989wd, Bhatia:2020itt, Hui:2001wy, Kamada:2022zwb},
\bea
\left<\sigma_{k\to 2} v_{rel}^{k-1}\right>_{\rm max}=\sum_{l} (2l+1)\frac{2^{\frac{3k-2}{2}}(\pi x)^{\frac{3k-5}{2}}}{g_{\rm DM}^{k-2}m_{\rm DM}^{3k-4}}\,.
\eea
and for non-identical particles in non-relativistic approximation with s-wave contribution \cite{Namjoo:2018oyn},
\bea
\left<\sigma_{k\to 2} v_{\rm rel}^{k-1}\right>~\leq ~2^{(3k-1)/2}(T/\pi)^{(5-3k)/2}S_k\frac{g_4 g_5}{g_1..g_k}\left(\frac{m_1+...+m_k}{m_1...m_k}\right)^{3/2}\,,
\eea
where $S_k$ is the symmetry factor associated with identical particles in the initial state, and $g_k$ counts the degrees of freedom of $i^{th}$ particle. In our case maximum value of $\rm 3{DM}\to 2{DM}$ annihilation are,
\bea
\left<\sigma_{\chi\chi\chi\to \chi\chi^*} v_{\rm rel}^{2}\right>~\leq \frac{48\sqrt{3}\pi^2}{m_{\chi}^3T^2}3!{\rm ~and~} \left<\sigma_{\chi\chi\chi^*\to \chi^*\chi^*} v_{\rm rel}^{2}\right>~\leq \frac{48\sqrt{3}\pi^2}{m_{\chi}^3T^2}2!\,.
\eea
However, these constraints are very mild for MeV scale DMs. The unitarity bound in the infinite scattering energy limit is given 
by \cite{Lerner:2009xg, Belanger:2012zr, Lee:1977eg, Horejsi:2005da, Hektor:2019ote},
\bea
|\lambda_{\chi H}|\leq8\pi\,,~|\lambda_{\chi}|\leq 4\pi\,.
\eea
The perturbative bound for this model is given by \cite{Bhattacharya:2019mmy, Choi:2021yps},
\bea
|\lambda_{\chi H}|\leq4\pi\,,~|\lambda_{\chi}|\leq\pi\,.
\eea
\paragraph{$\bullet$ Vacuum Stability\\}
The ncecessary condition to stabilise the potential (\ref{eq:lag}) is,
\bea
\lambda_H>0\,,~\lambda_{\chi}>0\,,~\lambda_{\chi H}+2\sqrt{\lambda_{\chi}\lambda_{H}}>0\,.
\eea
The maximal allowed value of the cubic parameter $\mu_3$ is approximately equal to $\mu_{3}|_{\rm max}\approx 2\sqrt{2}\sqrt{\lambda_{\chi}/\delta}~m_{\chi}$ where dimensioless parameter $\delta $ parameterises the energy difference of vacua \cite{Adams:1993zs}. For $\delta=2$, the $\mathbb{Z}_3$ breaking minimum is approximately degenerate with the SM minimum and gives the absolute stability bound \cite{Belanger:2012zr, Hektor:2019ote}.

\paragraph{$\bullet$ Kinetic equilibrium\\}
The dominating FO process for SIMP is $3_{\rm DM}\to 2_{\rm DM}$, which converts some fraction of DM (SIMP) mass into dark matter kinetic energy and gets 
distributed among themself and the scattering with SM bath particles transfer energy to the thermal bath, otherwise this will heat up the dark sector core and violates the 
astrophysical constraints derived from the structure formation \cite{Carlson:1992fn}. The kinetic equilibrium should be maintained between DM and SM bath 
minimally up to the freeze out of $\chi$ so that $\rm T_{SM}=T_{DM}$. The required condition to achieve this is 
$\langle\Gamma_{\rm \chi~SM\to \chi~SM}\rangle_{T=T_{\chi}^{\rm FO}}\gtrsim \mathcal{H}(T_{\chi}^{\rm FO})$ and is one the most important conditions 
for SIMP. DM can scatter with relativistic SM fermions, which is in thermal equilibrium during FO. As the relativistic fermions $(f)$ are abundant in the 
radiation-dominated era, the scattering process $\chi f \to \chi f$ is fast enough to keep dark and the visible sector in kinetic equilibrium during FO, respecting 
$\rm T_{\rm DM}=T_{\rm SM}$ \cite{Choi:2015bya, Choi:2017zww, Tulin:2017ara}.
We have estimated a bound on the portal coupling $(\lambda_{\chi H})$ for a sample of DM mass $(m_{\chi})$ in \autoref{app:D} to maintain kinetic equilibrium. 
Another way to achieve the kinetic equilibrium between two sectors is to extend the dark sector by introducing a relativistic new particle during FO \cite{Bernal:2015bla}. 
However, if the dark sector temperature is much less than the SM bath temperature, these two sectors can never reach kinetic equilibrium \cite{Bernal:2015ova, Bernal:2015xba}.
\paragraph{$\bullet$ Dark matter self scattering\\}
As SIMP type DM has sizeable self-interaction, which helps to solve the small-scale problems \cite{Tulin:2017ara, Spergel:1999mh} but is constraint by several observation constraints like,

$\star~$~Bullet~cluster \cite{Clowe:2003tk,Markevitch:2003at,Randall:2008ppe}:\quad $\frac{\sigma_{\rm self}}{m_{\rm DM}}\lesssim 1~ \rm cm^2gm^{-1}\,.$ 

$\star~$ Abell~cluster \cite{Kahlhoefer:2015vua}:\quad$1~ \rm cm^2gm^{-1}\lesssim\frac{\sigma_{\rm self}}{m_{\rm DM}}\lesssim 3~ \rm cm^2gm^{-1}\,.$

\noindent
A similar kind of bound is also available from cosmological simulation on self-interacting DM in galaxy cluster \cite{Peter:2012jh, Rocha:2012jg}.
\paragraph{$\bullet$ Dark Matter Mass limits\\}
BBN has a put significant constraint from different cosmological observations \cite{Nollett:2013pwa, Battaglieri:2017aum, Krnjaic:2019dzc}. 
So, it is an obvious choice for DM to freeze out before BBN; {\it i.e\,.~}$\rm T_{FO}>T_{BBN}\sim 0.1$ MeV gives an essential constraint on the 
masses of thermal DM. For DM masses above 10 MeV, BBN constitutes an important probe for the annihilation cross section of DM \cite{Depta:2019lbe, Giovanetti:2021izc}.
\paragraph{$\bullet$ Boltzmann Equation\\}
\noindent 
To get the present number density of SIMP DM, we have to solve the BEQ,
\bea
 \frac{dY_{\chi}}{dx}=-\frac{s}{x~\mathcal{H}(x)}\left[\langle\sigma v\rangle_{\chi\chi^*\to \rm SM ~SM}(Y_{\chi}^2-Y^{eq^2}_{\chi})+s\langle\sigma v^2\rangle_{3\chi\to 2\chi}(Y_{\chi}^3-Y_{\chi}^2Y^{eq}_{\chi})\right]\,,
 \label{eq:simp_beq}
\eea
where $x=m_{\chi}/T,~\mathcal{H}(x)\,=\,1.67\sqrt{g_*^{\rho}}m_{\chi}^2/\left(x^2\rm M_{pl}\right),$ $ {\bf s}  \,=\,\frac{ 2\pi^2}{45}g_{*}^{\bf s}\left(m_{\chi}/x\right)^3$ and $g_{*}^{ {\bf s} (\rho)}$ is the entropy (matter) degrees of freedom. The total DM yield is twice $Y_{\chi}$ as its conjugate particle ($\chi^*$) also has the same yield. Also note the presence of second term inside the parenthesis, absent in the WIMP, which dominates the SIMP like freeze out. A semi-analytical solution of eq\,.~\ref{eq:simp_beq} has been shown in \autoref{app:A} and a comparison with the numerical solution to match closely. 

\begin{figure}[htb!]
\centering
\subfloat[]{\includegraphics[width=0.475\linewidth]{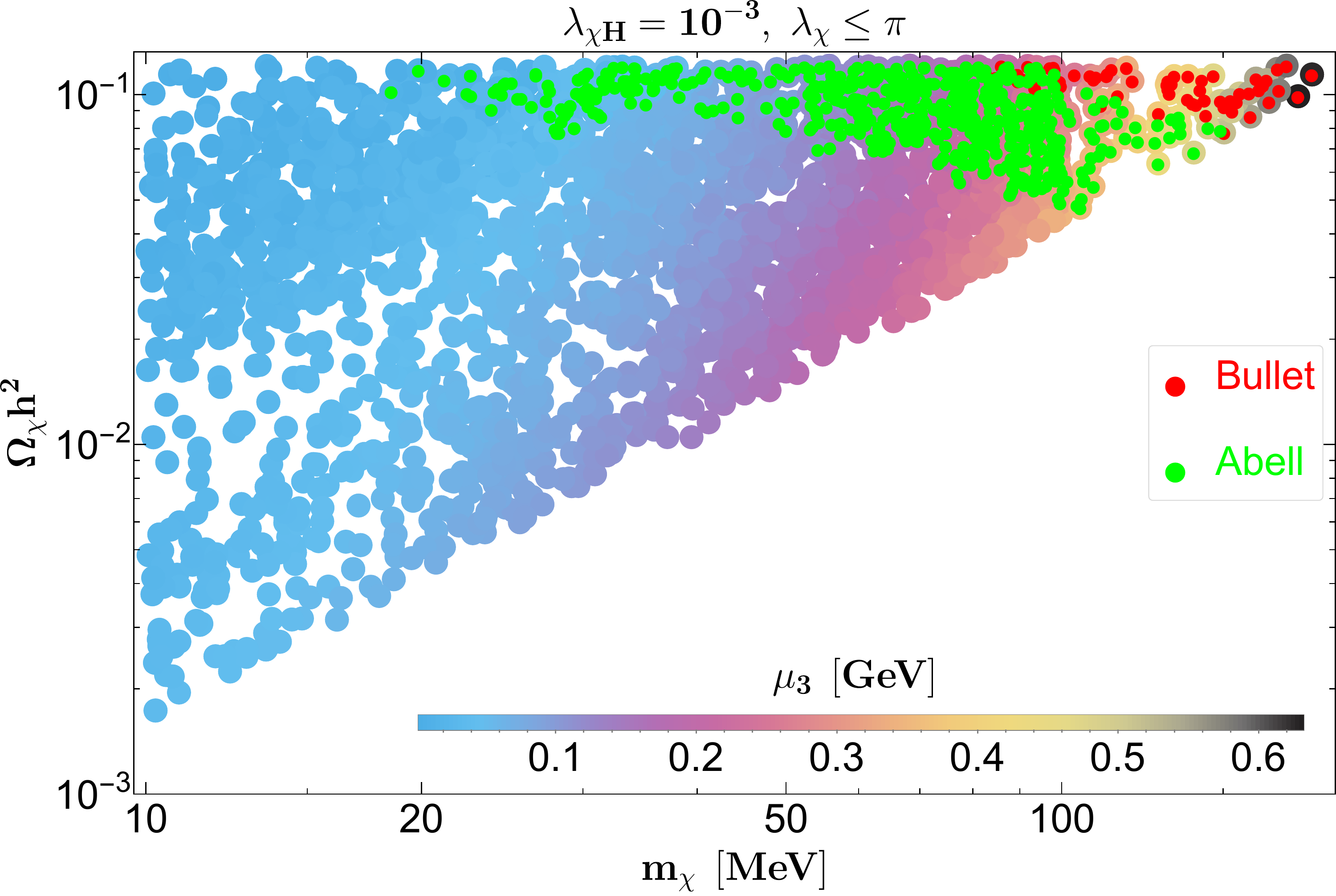}\label{fig:-mchi_mu3}}\quad
\subfloat[]{\includegraphics[width=0.475\linewidth]{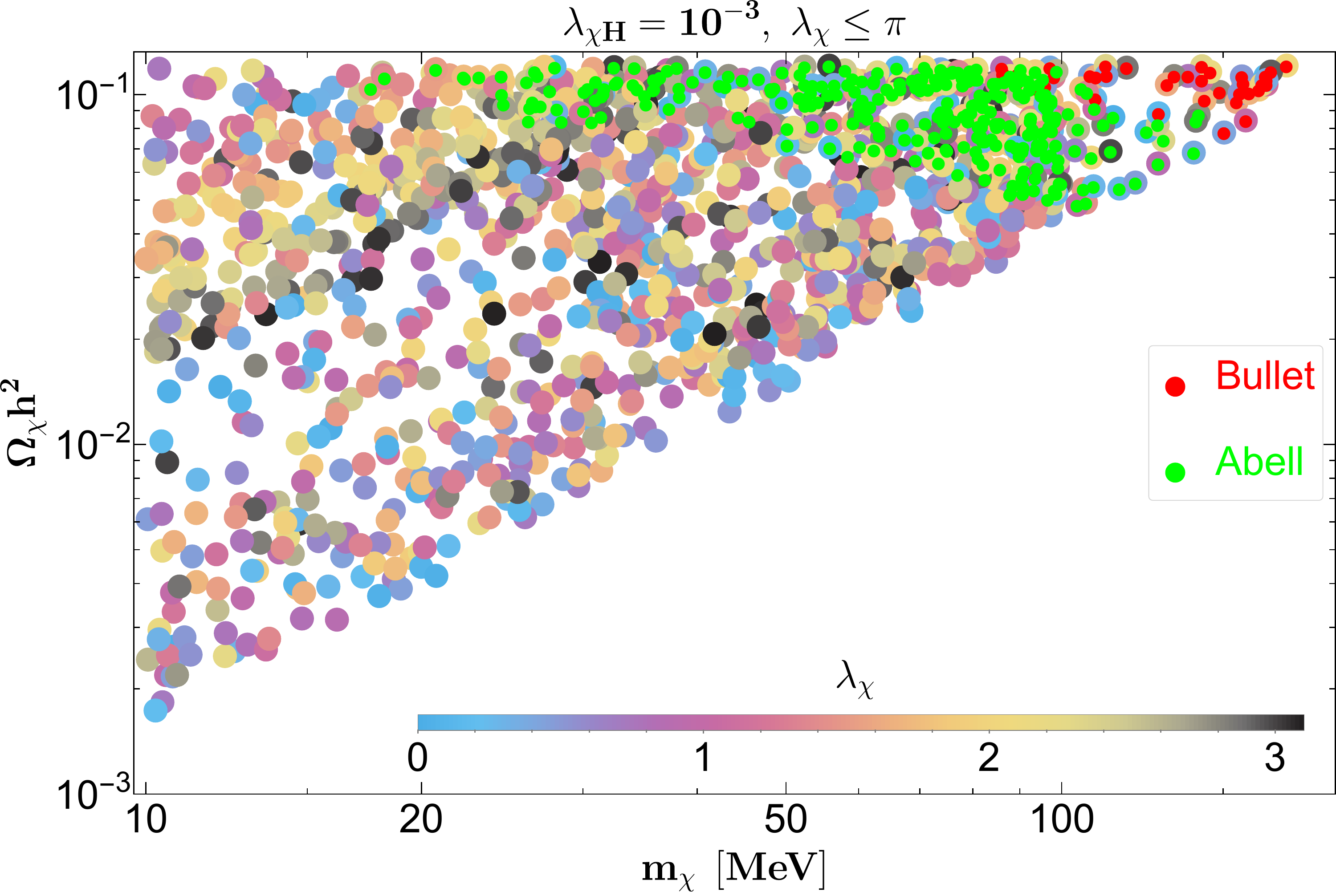}\label{fig:-mchi_lchi4}}
\caption{figs\,.~\ref{fig:-mchi_mu3} and \ref{fig:-mchi_lchi4} represent the relic and unitarity allowed parameter space in $\rm m_{\chi}-\Omega_{\chi}h^2$ plane for a 
complex scalar SIMP ($\chi$) described by the Lagrangian in eq. \eqref{eq:lag}. The variation of the relevant couplings ($\mu_3,\lambda_\chi$) is shown in the color bar, 
while the ones fixed is mentioned in the figure heading. Red and green points satisfy the self interaction limits from Bullet and Abell cluster bounds, respectively.}
\label{fig:simp_scan}
\end{figure}
The parameter space scan of the numerical solution of eq\,.~\ref{eq:simp_beq} for a complex scalar SIMP ($\chi$) described by the Lagrangian in 
eq\,.~\eqref{eq:lag} is shown in fig\,.~\ref{fig:simp_scan}. In figs\,.~\ref{fig:-mchi_mu3} and \ref{fig:-mchi_lchi4}, we show the relic under abundant parameter space in $\rm m_{\chi}-\Omega_{\chi}h^2$ planes and the color bar represents variation in cubic parameter $\mu_3$ (left) and quadratic coupling $\lambda_{\chi}$ (right) within appropriate limits. 
The SIMP relic density is inversely proportional to the DM mass and proportional to $\mu_3$ and $\lambda_{\chi}$ parameters, which is also reflected in 
figs\,.~\ref{fig:-mchi_mu3} and \ref{fig:-mchi_lchi4} respectively. With the enhancement of SIMP mass, DM self-annihilation cross section decreases, and the 
relic density is enhanced. To adjust the relic density, we have to increase the couplings within the theoretical limits. We are getting DM mass up to 
$\rm m_{\chi}\lesssim 200$ MeV in a single-component complex scalar SIMP scenario to acquire correct relic density. The self-scattering cross-section over DM mass, eq\,.~\ref{eq:simp_dm-self} strongly constrains the relic density allowed parameter space. After considering all of the available theoretical and cosmological constraints, 
few points respect the presently available DM self-interaction bound from Bullet and Abell cluster, marked by red and green points respectively as shown in fig\,.~\ref{fig:simp_scan}. This preliminary analysis helps us addressing the phenomenological distinction with the two component scenario considered later in this paper.
Before concluding this section, let us furnish some of the two component DM frameworks in combination with SIMP in tab\,.~\ref{tab:simp-scenrio} 
and remind that we focus on the pFIMP-SIMP combination, which was briefly mentioned in \cite{Bhattacharya:2022dco}.
\begin{table}[htb!]
\begin{center}\footnotesize
\begin{tabular}{|>\ccg{c}>\ccw{c}>\ccg{c}>\ccw{c}>\ccg{c}>\ccw{c}|}
\hline \rowcolor{red}Scenario&$\rm w-s$ & $\Gamma_{\rm SM ~SM\to w~w}$ & $ \Gamma_{\rm SM~SM\to s~s}$  & $ \Gamma_{\rm s~s\to w~w}$ & Self Annihilation \\\hline
I&FIMP-SIMP~~~~~~  & Feeble & Weak & Feeble&Weak-Strong\\
\rowcolor{lightgray}
II&pFIMP-SIMP~[this work] & Feeble & Weak & Weak &Weak-Strong\\ 
III&WIMP-SIMP ~~~~~~ & Weak & Weak & Weak/Feeble &Weak-Strong\\
IV&~SIMP-SIMP~\cite{Ho:2022erb,Choi:2021yps} & Weak & Weak & Weak/Feeble &Strong-Strong\\\hline
\end{tabular}
\end{center}
\caption{Generic two-component DM scenarios in the presence of a SIMP where w and s stand for two different or the same kind of dark sector particles, 
and $\Gamma_{i~j\to k~l}$ indicate the annihilation or production rate of the respective particle.}
\label{tab:simp-scenrio}
\end{table}

\section{pFIMP in presence of a SIMP: model independent analysis}
\label{sec:3}
Let us briefly recapitulate the essence of pFIMP \cite{Bhattacharya:2022dco, Bhattacharya:2022vxm} first. 
pFIMP is a kind of thermal DM, which has feeble interaction with the visible sector, but remains in equilibrium via interaction with the thermal DM partner. 
pFIMP therefore, can only be realised in a multicomponent DM scenario, when its partner is in thermal equilibrium. pFIMP freeze out therefore heavily depends 
on the partner's character, and the partner's detectability decides pFIMP's detection possibility \cite{Bhattacharya:2022vxm}.

The existence of pFIMP was demonstrated in presence of WIMP in  \cite{Bhattacharya:2022dco, Bhattacharya:2022vxm}. Here we show that SIMP being another kind of thermal 
DM, can accommodate a pFIMP DM in a multicomponent set up (the second possibility in tab\,.~\ref{tab:simp-scenrio}), with substantial interactions between them. 
The analysis has been done in two steps: firstly, a model-independent analysis on the pFIMP-SIMP scenario, in sec\,.~\ref{sec:3}, showing 
that the pFIMP dynamics is possible in the presence of a SIMP.  Secondly, we do the same exercise for the simplest two-component real (pFIMP) and complex scalar SIMP 
DM model, in sec\,.~\ref{mod:-simp-pfimp}.
In terms of yield $Y=n/\textbf{s}$ the two component cBEQ becomes,
\begin{gather}
\nonumber\frac{dY_{\rm w}}{dx}=\frac{2~\bf{s}}{x~\mathcal{H}(x)}\bigg[\frac{1}{\textbf{s}} \left(Y_{\rm SM}^{\rm eq }-Y_{\rm SM}^{\rm eq}\frac{Y_{\rm w}^2}{Y_{\rm w}^{\rm eq^2}}\right)\langle\Gamma\rangle_{\rm SM\to\rm w~w}\,+\,\Bigl(Y_{\rm SM}^{\rm eq^2}-Y_{\rm SM}^{\rm eq^2}\frac{Y_{\rm w}^2}{Y_{\rm w}^{\rm eq^2}}\Bigr)\langle\sigma v\rangle_{\rm{SM~ SM}\to\rm w ~w}\\+\Bigl(Y_{\rm s}^2-Y_{\rm s}^{\rm eq^2}\frac{Y_{\rm w}^2}{Y_{\rm w}^{\rm eq^2}}\Bigr)\langle\sigma v\rangle_{\rm s~s\to w~w}\bigg]\,,\label{eq:cbeq-miw}
\end{gather}
\begin{gather}
\nonumber\frac{dY_{\rm s}}{d x}=-\frac{\bf{s}}{x~\mathcal{H}(x)}\bigg[\left(Y_{\rm s}^2-Y_{\rm s}^{\rm eq^2}\right)\langle\sigma v\rangle_{\rm s~s\to SM~SM}\,+\,\textbf{s}\left(Y_{\rm s}^3-Y_{\rm s}^2Y_{\rm s}^{\rm eq}\right)\langle\sigma v^2\rangle_{3s\to 2s}\\+\left(Y_{\rm s}^2-Y_{\rm s}^{\rm eq^2}\frac{Y_{\rm w}^2}{Y_{\rm w}^{\rm eq^2}}\right)\langle\sigma v\rangle_{\rm s~s\to w~w}\bigg]\,.
\label{eq:cbeq-mis}
\end{gather}
In the above, $Y_{\rm w}$ refers to pFIMP yield, and $Y_{\rm s}$ defines SIMP yield. $\mathcal{H}(x)\,=\,1.67\sqrt{g_*^{\rho}}\mu_{\rm sw}^2/\left(x^2\rm M_{pl}\right)$, $\bf{s}$$\,=\,\frac{ 2\pi^2}{45}g_{*}^{\rm \bf{s}}\left(\mu_{\rm sw}/x\right)^3$$,~ Y_{ i}^{\rm eq}\,=\,\frac{45}{4\pi^4}\frac{g_i}{g_*^{\bf s}}\left(\frac{m_i}{\mu_{\rm sw}}x\right)^2K_2\left(\frac{m_i}{\mu_{\rm sw}}x\right)$. 
\begin{figure}[htb!]
\centering
\subfloat[]{\includegraphics[width=0.475\linewidth]{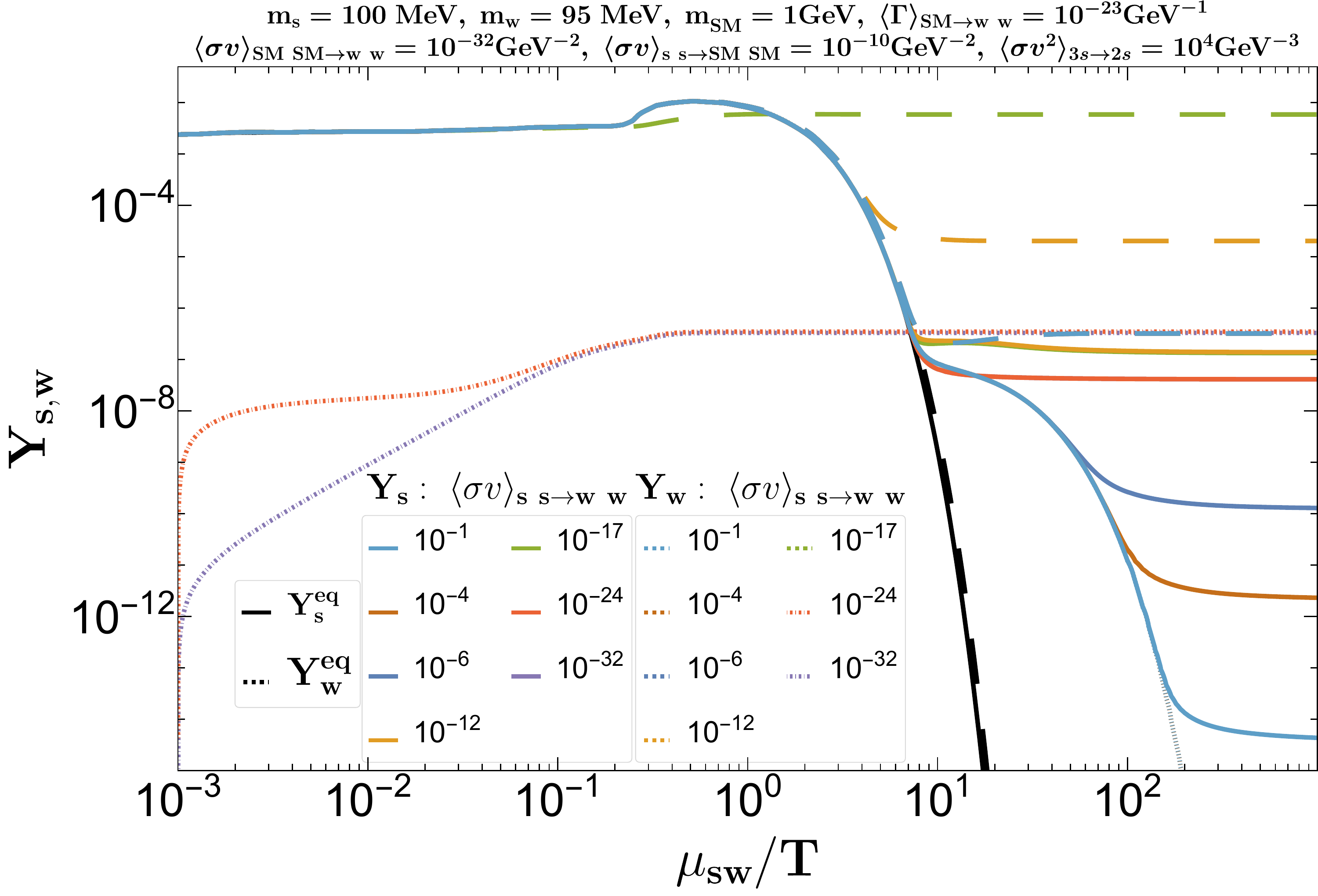}\label{fig:-yieldsgw_mi}}\quad
\subfloat[]{\includegraphics[width=0.475\linewidth]{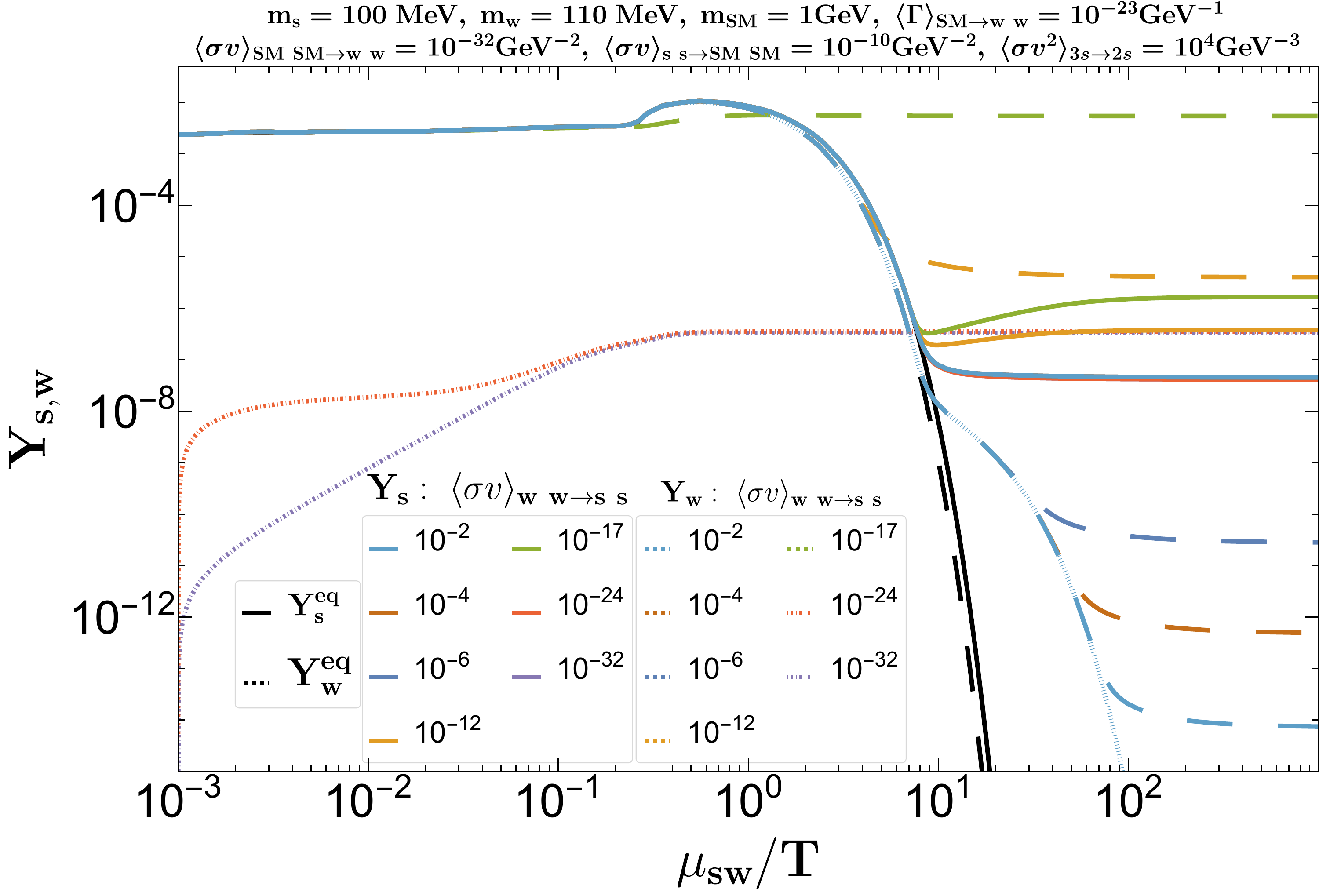}\label{fig:-yieldslw_mi}}

\subfloat[]{\includegraphics[width=0.475\linewidth]{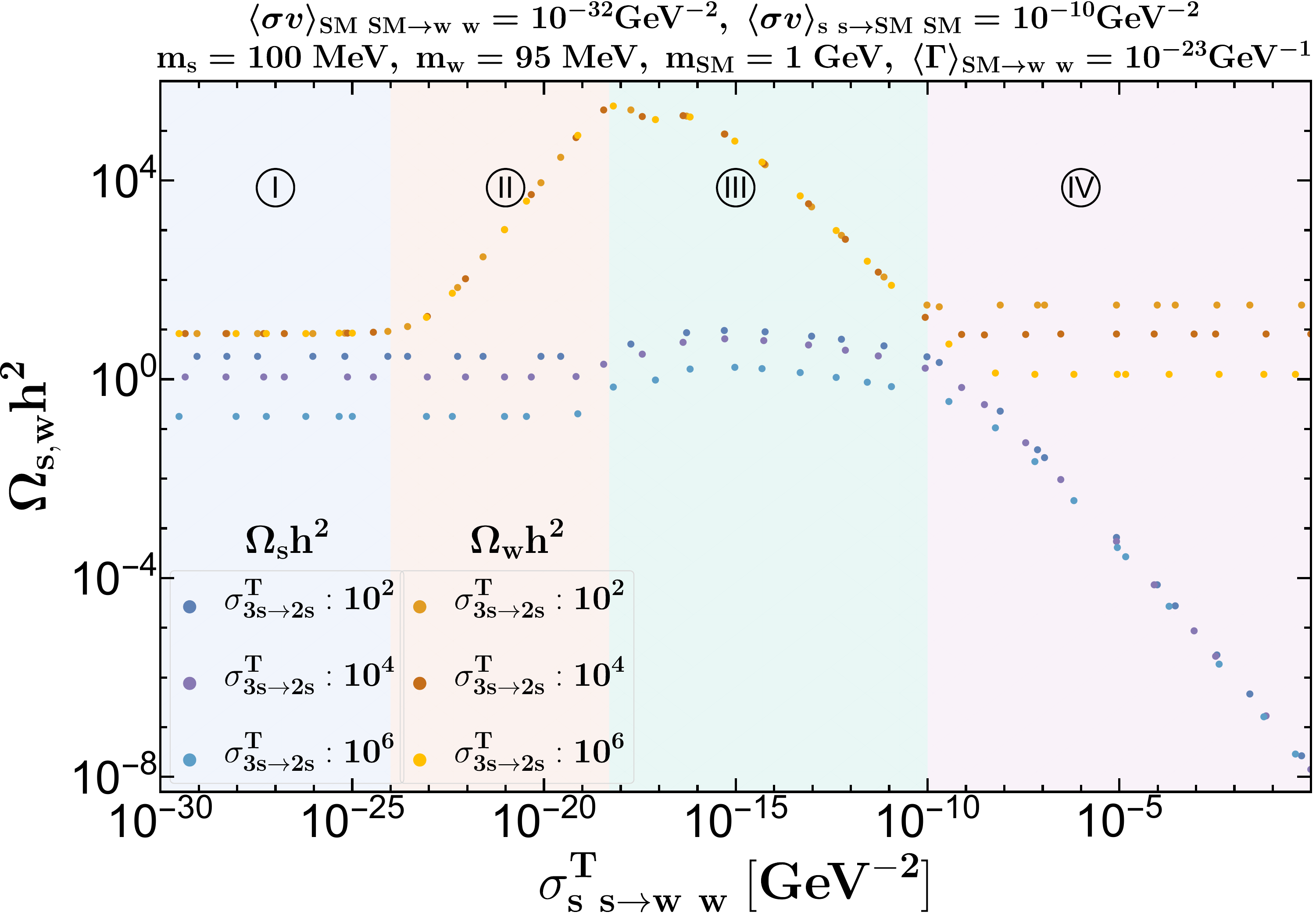}\label{fig:-relic_3s2s-ssww}}\quad
\subfloat[]{\includegraphics[width=0.475\linewidth]{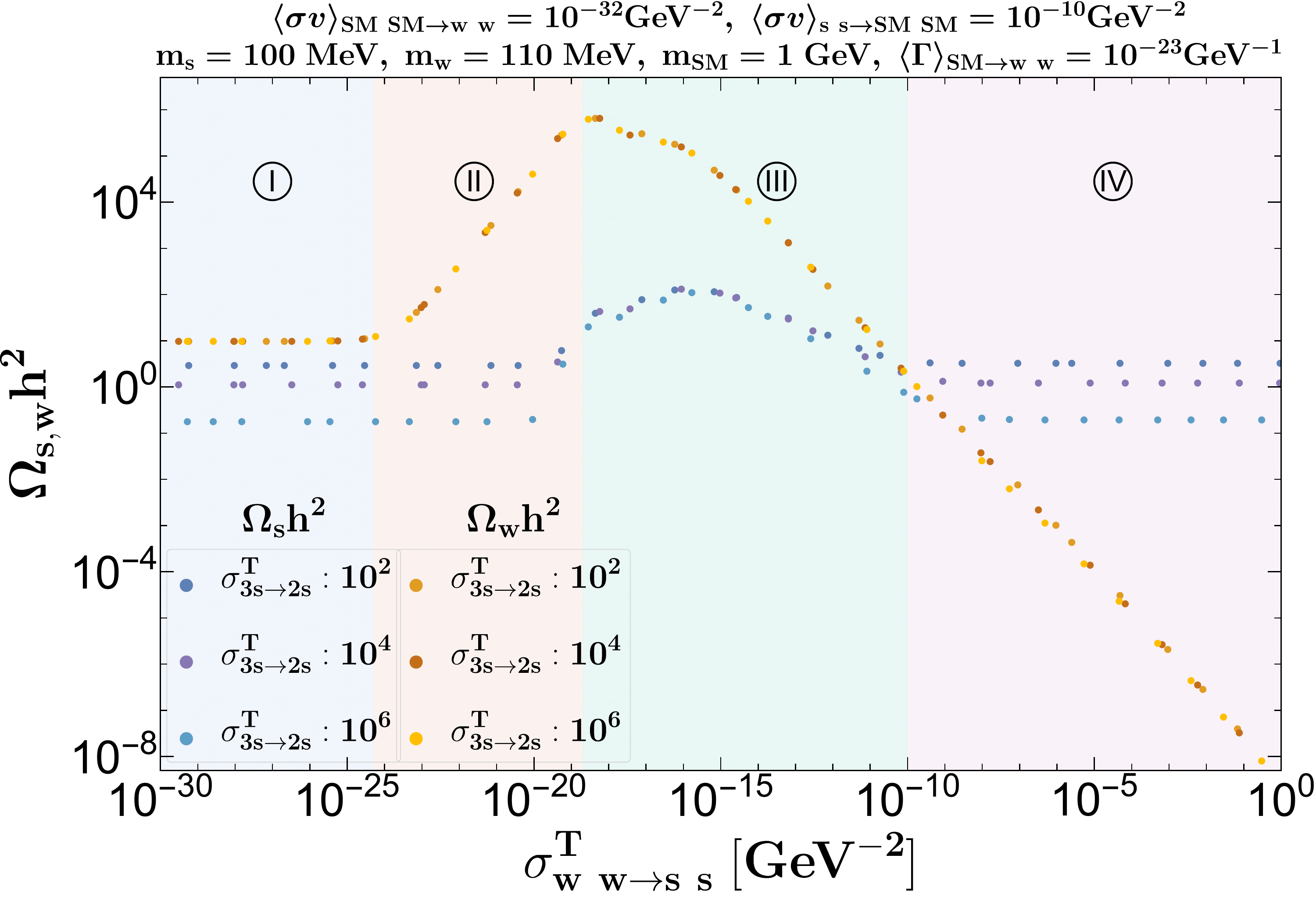}\label{fig:-relic_3s2s-wwss}}
\caption{Solution to cBEQs \ref{eq:cbeq-miw}, and \ref{eq:cbeq-mis} for pFIMP-SIMP scenario; figs\,.~\ref{fig:-yieldsgw_mi}, \ref{fig:-relic_3s2s-ssww} represent $\rm{m_s> m_w}$ case, while figs\,.~\ref{fig:-yieldslw_mi}, \ref{fig:-relic_3s2s-wwss} represent $\rm{m_s< m_w}$ case. Figs\,.~\ref{fig:-yieldsgw_mi}, \ref{fig:-yieldslw_mi} represent the variation of DM yield with dimensionless parameter $\mu_{\rm sw}/T$ for different values of conversion cross-section illustrated by different colored lines, solid for SIMP, dashed for pFIMP. Figs\,.~\ref{fig:-relic_3s2s-ssww}, \ref{fig:-relic_3s2s-wwss} represent 
DM relics as a function of DM-DM conversion cross-sections. Different colored dotted lines correspond to different SIMP annihilation cross-sections as mentioned 
in the plot legends.}
\label{fig:-sol-cbeq}
\end{figure}\par
The solution of cBEQ, ~\ref{eq:cbeq-miw} and \ref{eq:cbeq-mis}, is shown in figs\,.~\ref{fig:-yieldsgw_mi}, \ref{fig:-relic_3s2s-ssww} for $\rm{ m_w>m}_s$ and in figs\,.~\ref{fig:-yieldslw_mi}, \ref{fig:-relic_3s2s-wwss} for $\rm{ m_w<m}_s$. In the top panel of fig\,.~\ref{fig:-sol-cbeq}, we have varied the conversion cross-section, shown by different color thick (SIMP) and dotted/dashed (pFIMP) lines, while others $ m_s,$ $\rm m_w,$ $\langle\sigma v\rangle_{{\rm s~s\to SM~SM}},$ $\langle\sigma v\rangle_{{\rm w~w\to SM~SM}}$ and $\langle\sigma v^2\rangle_{\rm {3~s\to 2~s}}$ values are kept fixed as mentioned in the figure inset.
In this model-independent analysis, we have used the numerical values for thermal average cross-sections instead of a temperature-dependent functional form. The DM self-interaction constraints from different cosmological observations is yet to be considered, we do the same for model-dependent discussion.

The simplest way to explain fig\,.~\ref{fig:-sol-cbeq} by dividing the whole scenario into four regions, \circi{I} $\gamma_{\rm sw}< \gamma_{\rm ww}$;~\circii{II} $\gamma_{\rm ww}\lesssim \gamma_{\rm sw}\ll\gamma_{3s\to 2s} $;~\circiii{III} $\gamma_{\rm ww}\ll \gamma_{\rm sw}<\gamma_{3s\to 2s} $;~\circiv{IV} $\gamma_{3s\to 2s}\lesssim \gamma_{\rm sw} $. We define, $\gamma_{\rm ww}=\langle\sigma v\rangle_{\rm sm~sm\to\rm w~w}n_{\rm sm}^2\equiv \langle\Gamma \rangle_{\rm sm\to\rm w~w}n_{\rm sm}$,~$\gamma_{\rm sw}=\langle\sigma v\rangle_{\rm s~s\to\rm w ~w}n_{\rm s}^2$ and $\gamma_{3s\to 2s}=\langle\sigma v^2\rangle_{\rm 3s\to 2s}n_{\rm s}^3=\langle\sigma v^2\rangle_{\rm 2s\to 3s}n_{\rm s}^2$.

Region \circi{I} shown by the light blue shaded region in figs\,.~\ref{fig:-relic_3s2s-ssww} and \ref{fig:-relic_3s2s-wwss}, FIMP number density $n_{\rm w}$ is independent of the 
$\sigma^T_{s~s\to\rm w~w}$ due to the negligible DM-DM conversion cross-section compared to DM production rate from SM particles. This is also visible in 
figs\,.~\ref{fig:-yieldsgw_mi} ($\rm m_s>m_w$) and \ref{fig:-yieldslw_mi} ($\rm m_s<m_w$) from blue dot-dashed lines. This scenario corresponds to a pure FIMP for 
$Y_{\rm w}$ and pure SIMP for $Y_{\rm s}$.

Region \circii{II} refers to light red regions in figs\,.~\ref{fig:-relic_3s2s-ssww}, \ref{fig:-relic_3s2s-wwss} and is also represented by 
red dot-dashed lines in figs\,.~\ref{fig:-yieldsgw_mi}, \ref{fig:-yieldslw_mi}. In this regime, $\rm\gamma_{ ww}\lesssim \gamma_{ sw}\ll\gamma_{3s\to 2s} $, 
SIMP number density remains unaffected due to small conversion rate compared to $\rm \sigma^T_{3s\to 2s}$ but FIMP number density $(\rm n_w)$ 
is enhanced with larger conversion from SIMP, as reflected in all the four figs. This limit still represents a combination of SIMP and FIMP DMs, where FIMP production 
is affected by conversion from SIMP, but is mostly overabundant.

Suppose we further enhance the DM-DM conversion rate. In that case, FIMP ($\rm w$) reaches thermal equilibrium and follows equilibrium before freeze out. 
At the same time, SIMP number density is slightly enhanced due to opening up a new annihilation or production channel, see region \circiii{III} in figs\,.~\ref{fig:-relic_3s2s-ssww}, \ref{fig:-relic_3s2s-wwss} and green, yellow dashed lines in figs\,.~\ref{fig:-yieldsgw_mi}, \ref{fig:-yieldslw_mi}. After equilibration, we call the FIMP ($\rm w$) 
as pFIMP as it has a feeble visible connection but is weakly connected with partner SIMP type DM and makes it a two-component SIMP-pFIMP scenario.
With further enhancement of conversion rate $(\gamma_{\rm sw})$, both SIMP and pFIMP number density decreases, but before freeze out, they follow a 
modified equilibrium distribution depending on the choice of mass hierarchy, as shown by the dotted yellow and green lines in fig\,.~\ref{fig:-yieldsgw_mi}. 
In this regime, for $\rm (m_s<m_{w})$ (in fig\,.~\ref{fig:-yieldslw_mi}) SIMP number density is suddenly enhanced after decoupling from the thermal bath 
due to the production from pFIMP, which is also seen in fig\,.~\ref{fig:-relic_3s2s-ssww}. 
The expression of modified equilibrium number densities that the particles acquire in fig\,.~\ref{fig:-yieldsgw_mi} and \ref{fig:-yieldslw_mi} for two 
hierarchies are given by eqs\,.~\ref{eq:-wimp_modify-eq} and \ref{eq:-simp_modify-eq} respectively,
\begin{align}
 n_{\rm w}^{\rm eq^{\prime}}= n_{\rm w}^{\rm eq}\dfrac{n_{s}}{n_{s}^{eq}}\,\quad \text{for} ~{\rm m_w}>m_s\,,
 \label{eq:-wimp_modify-eq}
\end{align}
{\small
\begin{align}
n_{ s}^{\rm eq^{\prime}}\,=\,n_{ s}^{\rm eq^{}}\left[\frac{\langle\sigma v\rangle_{ s ~s\to \rm SM~SM}+\left(\frac{n_{ s}^2}{n_{ s}^{\rm eq}}\right)\langle\sigma v^2\rangle_{ 3s\to 2s}+\left(\frac{n_{\rm w}}{n_{\rm w}^{\rm eq}}\right)^2\langle\sigma v\rangle_{ s~s\to\rm w~w}}{\langle\sigma v\rangle_{ s ~s\to\rm SM~SM}+n_{ s}\langle\sigma v^2\rangle_{ 3s\to 2s}+\langle\sigma v\rangle_{ s~s\to\rm w~w}}\right]^{1/2}\text{for} ~{{\rm m}_s}>\rm m_{w}\,.
\label{eq:-simp_modify-eq}
\end{align}}

Even larger enhancement of DM conversion compared to the self-annihilation of SIMP, $\rm\gamma_{3s\to 2s}\lesssim \gamma_{ sw} $, gives a completely 
different scenario, as shown in \circiv{IV} of figs\,.~\ref{fig:-relic_3s2s-ssww}, \ref{fig:-relic_3s2s-wwss} and by deep blue, brown, light blue lines in figs\,.~\ref{fig:-yieldsgw_mi}, \ref{fig:-yieldslw_mi}. The conversion rate depends on the conversion cross-section as well as the number density during or before the heavier component freeze out. 
When the conversion is larger than SIMP self-annihilation, the heavier component continues to deplete into the lighter ones even after decoupling from the thermal bath.
Since the lighter component has frozen out before, the heavier component goes into the modified equilibrium; see the light blue dashed line in figs\,.~\ref{fig:-yieldsgw_mi}, \ref{fig:-yieldslw_mi}. With the enhancement of the conversion cross-section, the number density for lighter one doesn't change much due to the gradual decrease of heavy particle 
number density with the enhancement of conversion cross-section, and both simultaneously fix the light particle number density in region \circiv{IV} of figs\,.~\ref{fig:-relic_3s2s-ssww}, \ref{fig:-relic_3s2s-wwss}. However, here the SIMP is not purely a SIMP as its freeze out is mostly governed by conversion. 

\section{A model example of pFIMP-SIMP}
\label{mod:-simp-pfimp}

Let us now discuss a model example of pFIMP-SIMP set up. Here we introduce a real scalar $\phi$ 
and a complex scalar singlet field $\chi$ as DM components, which are stable under 
$\mathbb{Z}_2\otimes\mathbb{Z}_3$ symmetry and charge assignments are shown in tab\,.~\ref{tab:tab1}. 
\begin{table}[htb!]
\begin{center}
\begin{tabular}{|c|c|c|}\hline
\rowcolor{magenta!15}\bf Dark Fields&  $\mathbb{Z}_2$&$ \mathbb{Z}_3$\\\hline
\rowcolor{cyan!15}$\chi$&$\chi$&$\omega\chi$ or $\omega^2\chi$\\
\rowcolor{lime!15}$\phi$&-$\phi$&$\phi$\\\hline
\end{tabular}
\end{center}
\caption{Dark sector fields ($\chi$ and $\phi$) and their quantum numbers where $\omega=e^{i2\pi/3}$.}
\label{tab:tab1}
\end{table} 
The dark sector Lagrangian obeying the symmetry can be written as,
\begin{align}
\nonumber\mathcal{L}_{}\,=\,&\mathcal{L}_{\rm SM}+\mu_H^2H^{\dagger}H-\lambda_H (H^{\dagger}H)^2+\frac{1}{2}|\partial_{\mu}\phi |^2-\frac{1}{2}\mu_{\phi}^2\phi^2-\frac{1}{4\,!}\lambda_{\phi}\phi^4+|\partial_{\mu}\chi |^2-\mu_{\chi}^2|\chi|^2
\\&-\lambda_{\chi}|\chi^*\chi|^2-\frac{1}{2}\mu_{3}(\chi^3+\chi^{*^3})-\frac{1}{2}\lambda_{\phi H}\phi^2H^{\dagger}H-\lambda_{\chi H}|\chi|^2H^{\dagger}H-\frac{1}{2}\lambda_{\chi\phi}|\chi|^2\phi^2\,.
\label{eq:model}
\end{align}
After spontaneous symmetry breaking, $H=\left(0~\frac{v+h}{\sqrt{2}}\right)^{T}$.
From the interactions as in Eq. \eqref{eq:model}, $\chi$ having $3\to 2$ depletion, can be a SIMP, while $\phi$ can be pFIMP. Understandably, this is the simplest 
model example of pFIMP-SIMP set up. We are interested in a small mass ($\sim$ MeV) scale of pFIMP, as the SIMP is validated in that mass range only, as discussed before. 
We choose small values of Higgs portal coupling $\lambda_{\phi H}$. This helps the pFIMP to remain out of thermal bath, at the same time, helps evading 
Higgs to invisible branching ratio. We will also choose the other Higgs portal coupling $\lambda_{\chi H}$ to be small as well to adhere to the SIMP limit of $\chi$. This 
also takes care of the invisible Higgs branching ratio to $\chi$ pair. 
\subsection{Solution of coupled Boltzmann Equation and relic density}
Assuming CP conservation existing inside the dark sector, the coupled Boltzmann Equation for two DMs, 
where $Y_s=Y_{\chi}+Y_{\chi^*}$ with $Y_{\chi}=Y_{\chi^*}$ is given by,
\begin{gather}
\nonumber\dfrac{dY_{\phi}}{dx}=\frac{2~\bf{s}}{x~\mathcal{H}(x)}\bigg[\frac{1}{\textbf{s}} \left(Y_{\rm h}^{\rm eq }-Y_{\rm h}^{\rm eq}\frac{Y_{\phi}^2}{Y_{\phi}^{\rm eq^2}}\right)\langle\Gamma\rangle_{\rm h\to \phi~\phi}\,+\,\Bigl(Y_{\rm SM}^{\rm eq^2}-Y_{\rm SM}^{\rm eq^2}\frac{Y_{\phi}^2}{Y_{\phi}^{\rm eq^2}}\Bigr)\langle\sigma v\rangle_{\rm{SM~ SM}\to\rm \phi ~\phi}\\+\dfrac{1}{4}\Bigl(Y_{\rm s}^2-Y_{\rm s}^{\rm eq^2}\frac{Y_{\rm \phi}^2}{Y_{\rm \phi}^{\rm eq^2}}\Bigr)\langle\sigma v\rangle_{\rm \chi~\chi^*\to \phi~\phi}\bigg]\,,
\end{gather}
\begin{gather}
\nonumber\dfrac{dY_{s}}{dx}=-\frac{\bf{s}}{x~\mathcal{H}(x)}\biggl[\frac{1}{2}(Y_{s}^2-Y_{s}^{\rm eq^2})\langle\sigma v\rangle_{\chi~\chi^*\to \rm SM~SM}+\frac{\bf{s}}{4}(Y_{s}^3-Y_{s}^2Y_{s}^{\rm eq})\langle\sigma v^2\rangle_{ 3\chi\to 2\chi}\\+\frac{1}{2}\left(Y_{s}^2-Y_{s}^{\rm eq^2}\frac{Y_{\phi}^2}{Y_{\phi}^{\rm eq^2}}\right)\langle\sigma v\rangle_{\chi~\chi^*\to\phi ~\phi}\biggr]\,.
\label{eq:cbeq}
\end{gather}
In the above, $\rm SM=\{h,~W^{\pm},~Z,~leptons,~quarks\}$, {\bf s} represents entropy density, $Y^{\rm eq}_i=n_{i}^{\rm eq}/{\bf s}$ is the equilibrium yield of 
$i^{th}$ particle has mentioned in previous section, and,
\begin{align}
\nonumber\langle\sigma v^2\rangle_{\rm 3\chi\to 2\chi}=2\left(\langle\sigma v^2\rangle_{\chi\chi\chi\to \chi\chi^*}+\langle\sigma v^2\rangle_{\chi\chi^*\chi^*\to \chi\chi}\right)\,.
\end{align}
The total relic density is written in terms of DM yields obtained from the solution of cBEQ~\ref{eq:cbeq}, after freeze out, 
\bea
{\rm \Omega_{\rm DM}h^2} = 2.74385 \times 10^8 \left[m_{\chi}Y_{s}+m_{\phi}Y_{\phi}\right]_{x\to \infty}\,.
\eea

\begin{figure}[htb!]
\centering
\subfloat[]{\includegraphics[width=0.475\linewidth]{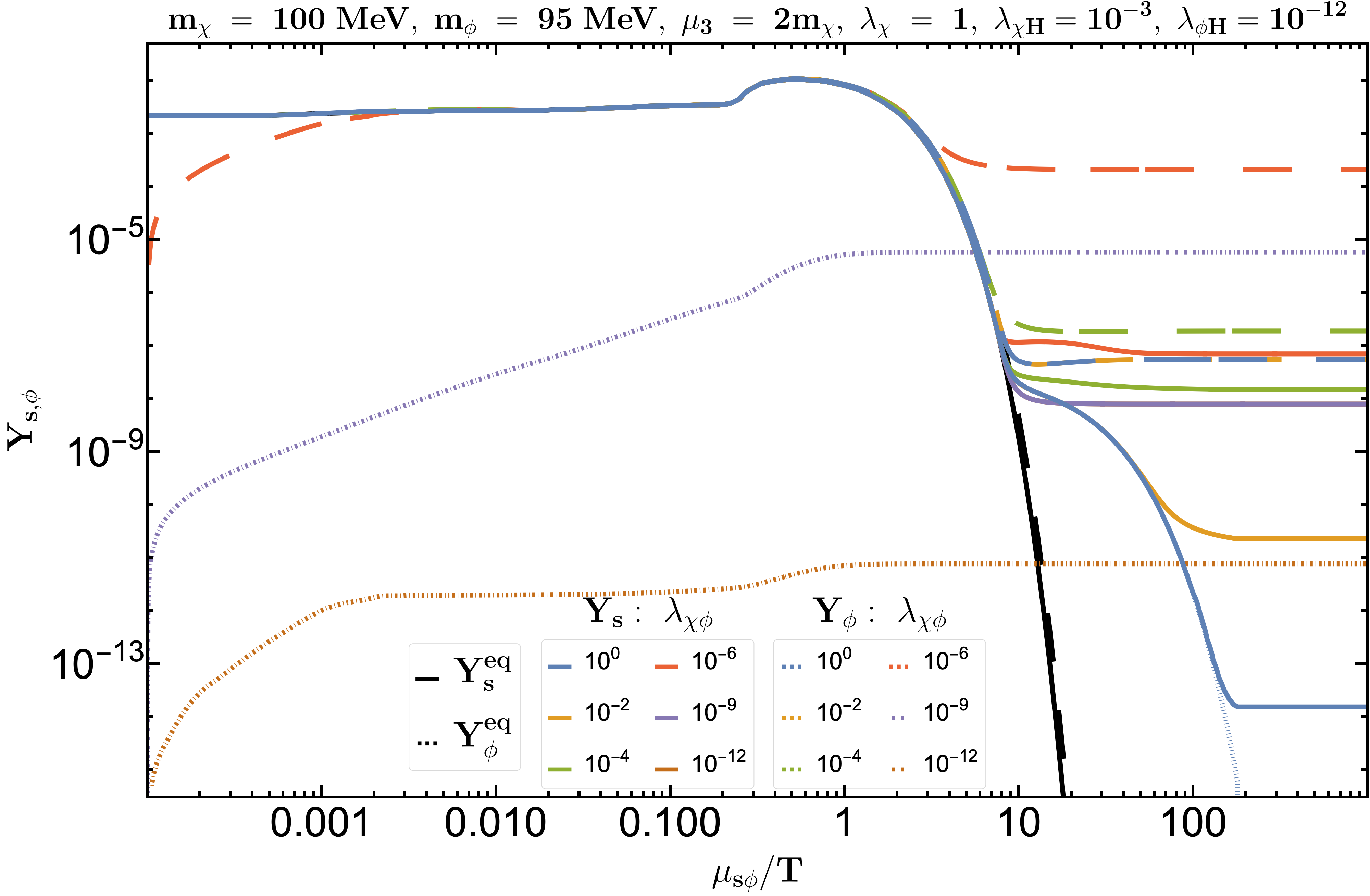}\label{fig:-yieldmchigmphi}}\quad
\subfloat[]{\includegraphics[width=0.475\linewidth]{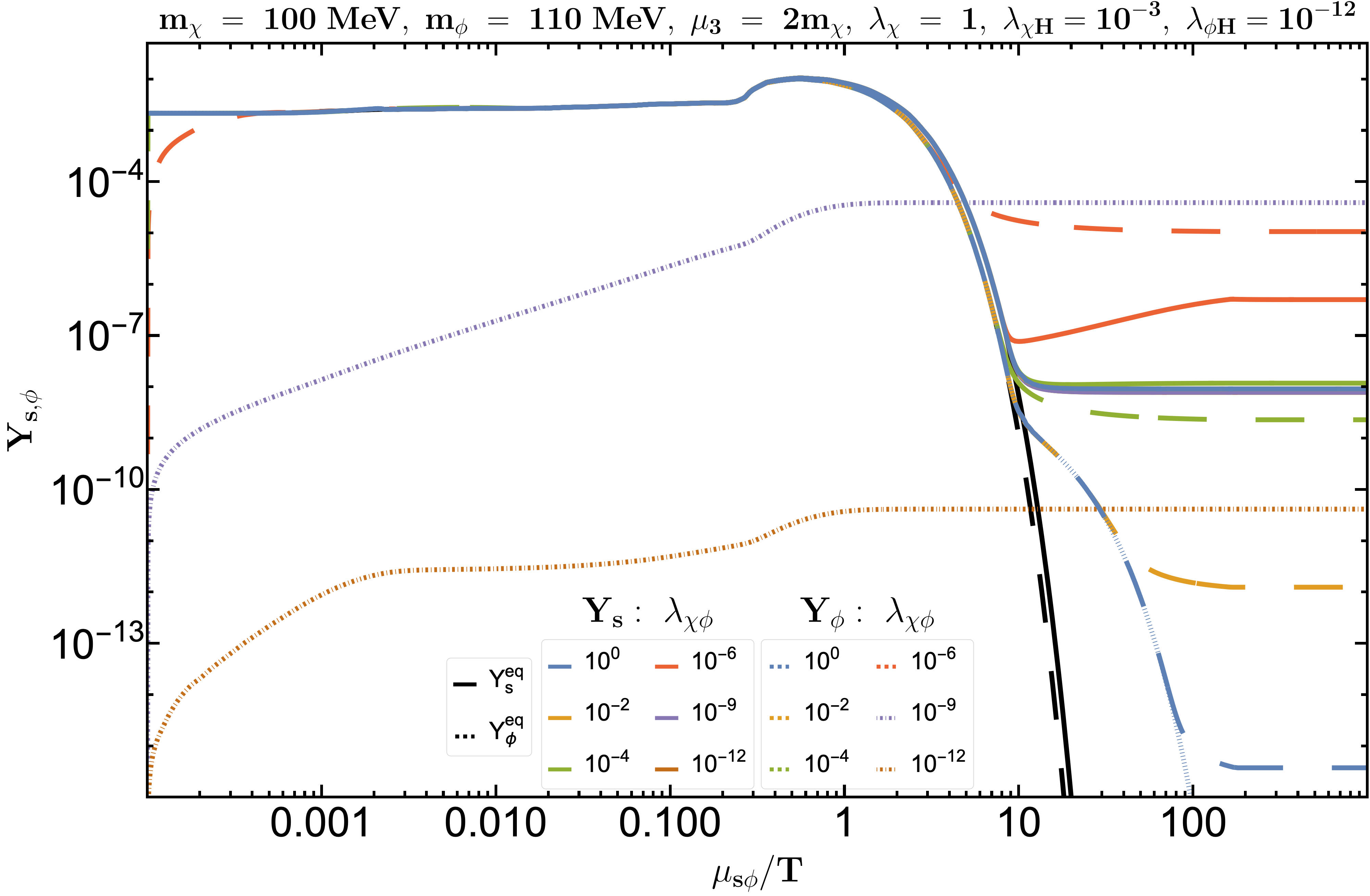}\label{fig:-yieldmchilmphi}}
\caption{Solution to the cBEQ~\ref{eq:cbeq}, where figs\,.~\ref{fig:-yieldmchigmphi} and \ref{fig:-yieldmchilmphi} represents $m_{\chi}>m_{\phi}$ and $m_{\chi}<m_{\phi}$ scenarios respectively. The thick, dashed and dotted color lines represent the SIMP, FIMP and pFIMP cases respectively. Different colors show $\lambda_{\chi\phi}$ variation 
as mentioned in the figure inset. The thick black and dashed lines represent the SIMP and pFIMP equilibrium yields. Other parameter kept fixed are mentioned in the figure heading.}
\label{fig:-yieldmchiglmphi}
\end{figure}

The solution of cBEQ~\ref{eq:cbeq} provides the yield of SIMP $(Y_{\chi})$ and pFIMP $(Y_{\phi})$ as shown in fig\,.~\ref{fig:-yieldmchiglmphi} 
with the variation of $\mu_{s\phi}/T$ for different mass hierarchies; fig\,.~\ref{fig:-yieldmchigmphi} shows the case 
for $m_{\chi}>m_{\phi}$ and fig.~\ref{fig:-yieldmchilmphi} represents the case for $m_{\chi}<m_{\phi}$.
The explanation of this figure is similar to that of fig\,.~\ref{fig:-sol-cbeq} (top row). The only change here is that we vary the free-parameter $\lambda_{\chi \phi}$ 
in solving cBEQ instead of choosing a numerical value for the conversion cross-section $\rm\sigma^T_{ ss\to ww}$ as done before. 
For small values of $\lambda_{\chi\phi}$ $(10^{-12},~10^{-9})$, $\phi$ shows pure FIMP like behaviour before freezing-in, as 
shown by the violet and orange dot-dashed lines in figs for both the mass hierarchies.
Enhancement of $\lambda_{\chi\phi}$ $(\sim 10^{-6})$ makes $\phi$ reach thermal bath and follow the equilibrium distribution before freezing out; 
$\phi$ becomes pFIMP here and is represented by dashed red lines in both the figs. At this moment, 
the number density of SIMP $\chi$ is enhanced due to new production channel opening up via conversion.
With further enhancement of $\lambda_{\chi\phi}$ $(\gtrsim 10^{-6}-10^{-4})$, we observe a decrease in both SIMP and pFIMP number densities, shown 
by green thick and dashed lines in fig\,.~\ref{fig:-yieldmchiglmphi} for both the mass hierarchies. This serve as the ideal pFIMP-SIMP parameter space, where 
we achieve under abundance for both the DM components. If $\lambda_{\chi\phi}$ is enhanced further to $\sim \{10^{-2},10^{0}\}$ then 
the annihilation of the heavier DM component to the lighter one is possible even after the freeze out of the lighter component, then the 
heavier component follows modified equilibrium distribution following eqs\,.~\ref{eq:-simp_modify-eq}, and \ref{eq:-wimp_modify-eq} before 
freezing out. The yellow $(\lambda_{\chi\phi}\sim10^{-2})$ and light blue $(\lambda_{\chi\phi}\sim10^{0})$ thick (SIMP)
and dashed (pFIMP) lines represent DM yields following such modified distributions. 
\subsection{Constraints on the model parameter space}

\begin{figure}[htb!]
\centering
\subfloat[]{\includegraphics[width=0.475\linewidth]{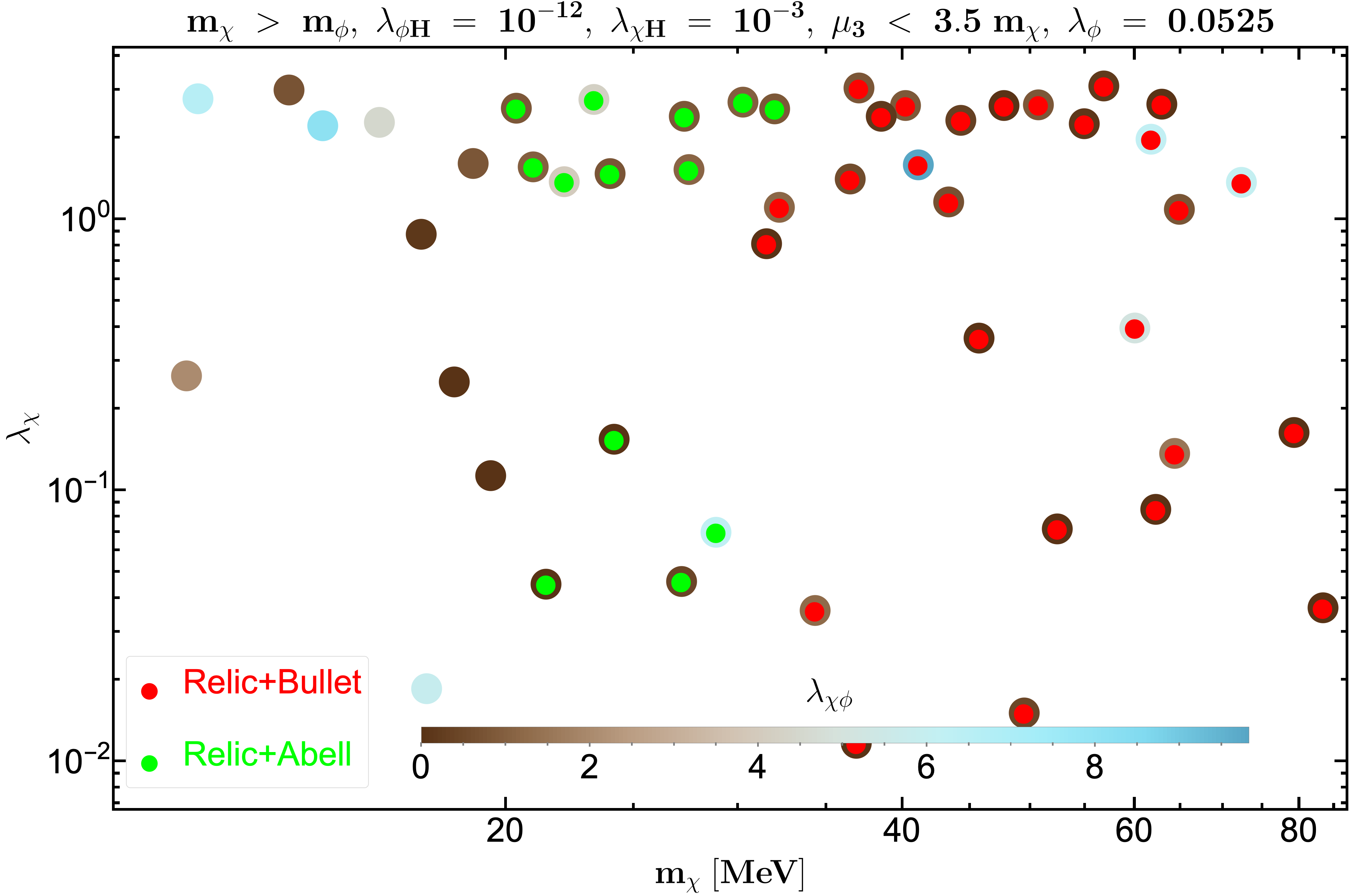}\label{fig:-pfimp-simp_chigphi-mchi-lchi-lchiphi}}\quad
\subfloat[]{\includegraphics[width=0.475\linewidth]{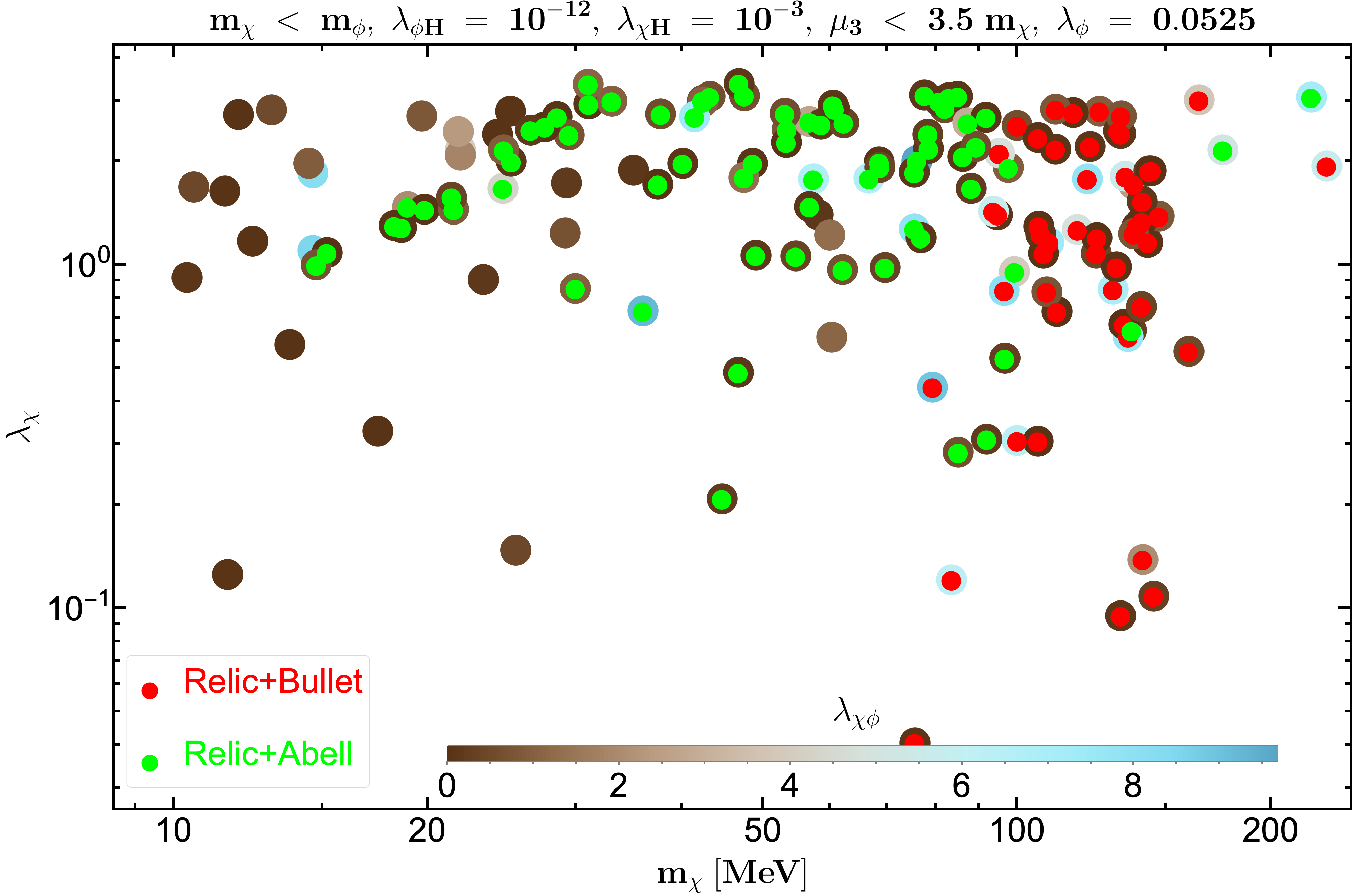}\label{fig:-pfimp-simp_chilphi-mchi-lchi-lchiphi}}

\subfloat[]{\includegraphics[width=0.475\linewidth]{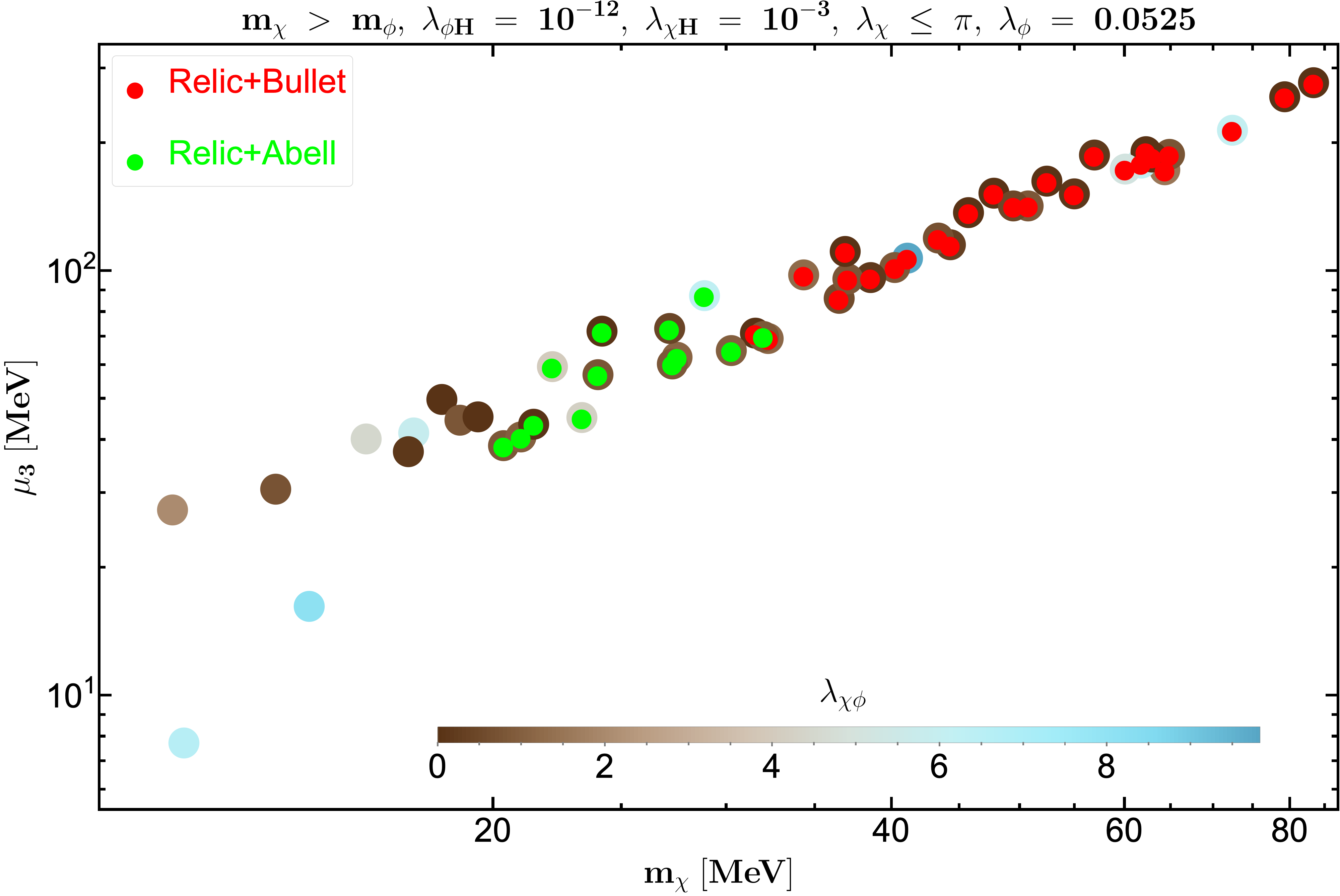}\label{fig:-pfimp-simp_chigphi-mchi-mu3-lchiphi}}\quad
\subfloat[]{\includegraphics[width=0.475\linewidth]{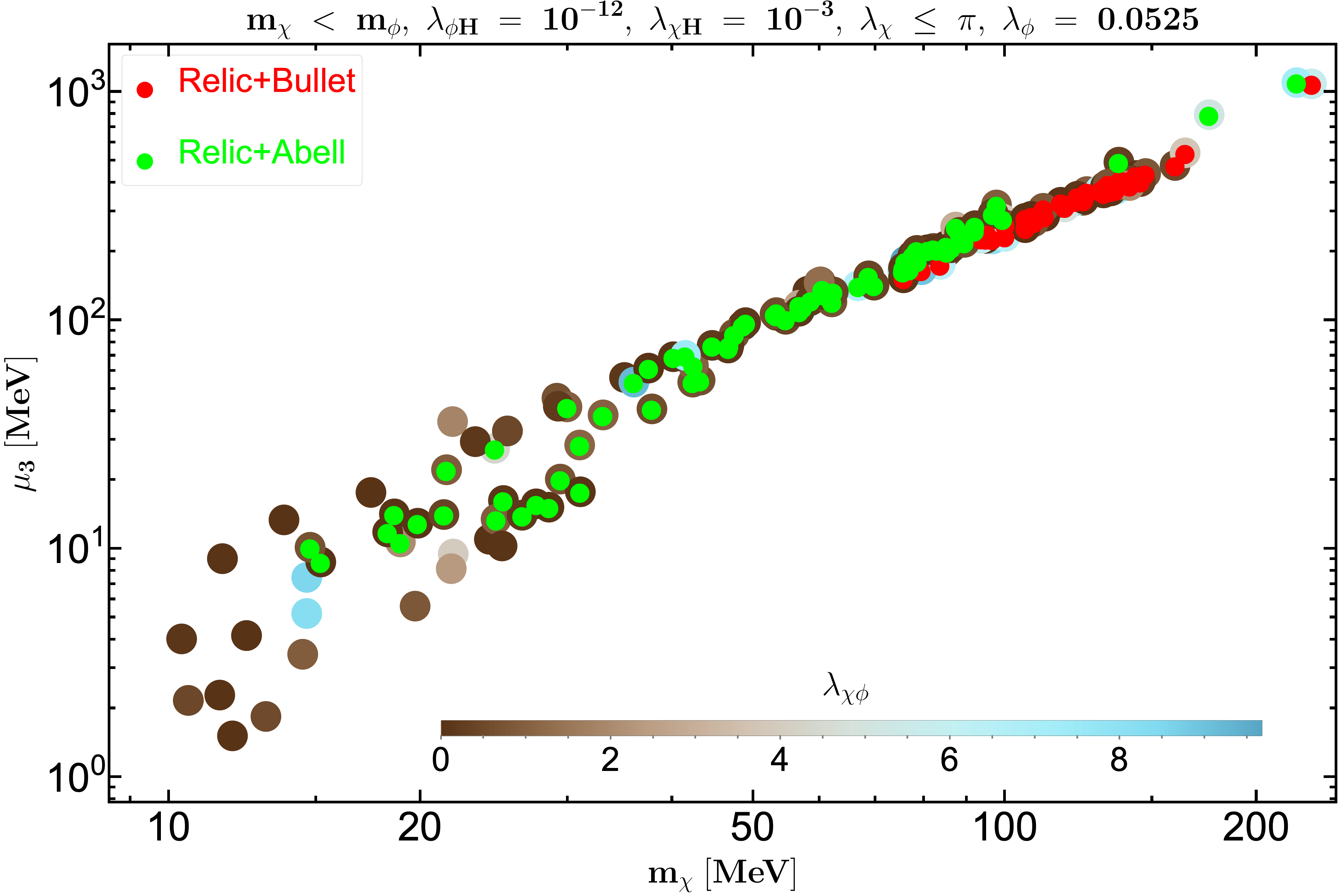}\label{fig:-pfimp-simp_chilphi-mchi-mu3-lchiphi}}

\subfloat[]{\includegraphics[width=0.475\linewidth]{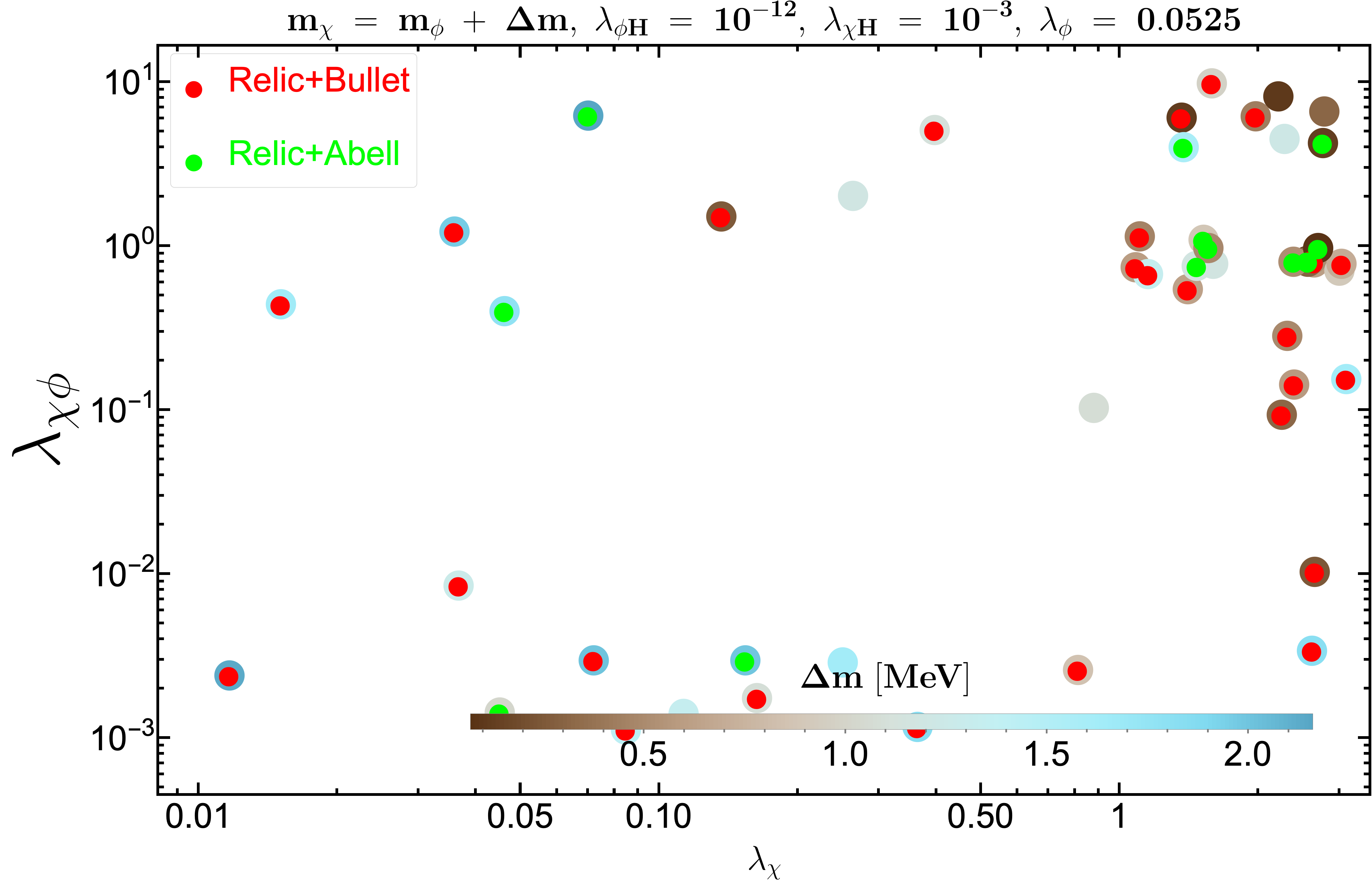}\label{fig:-pfimp-simp_chigphi-lchi-lchiphi-dm}}\quad
\subfloat[]{\includegraphics[width=0.475\linewidth]{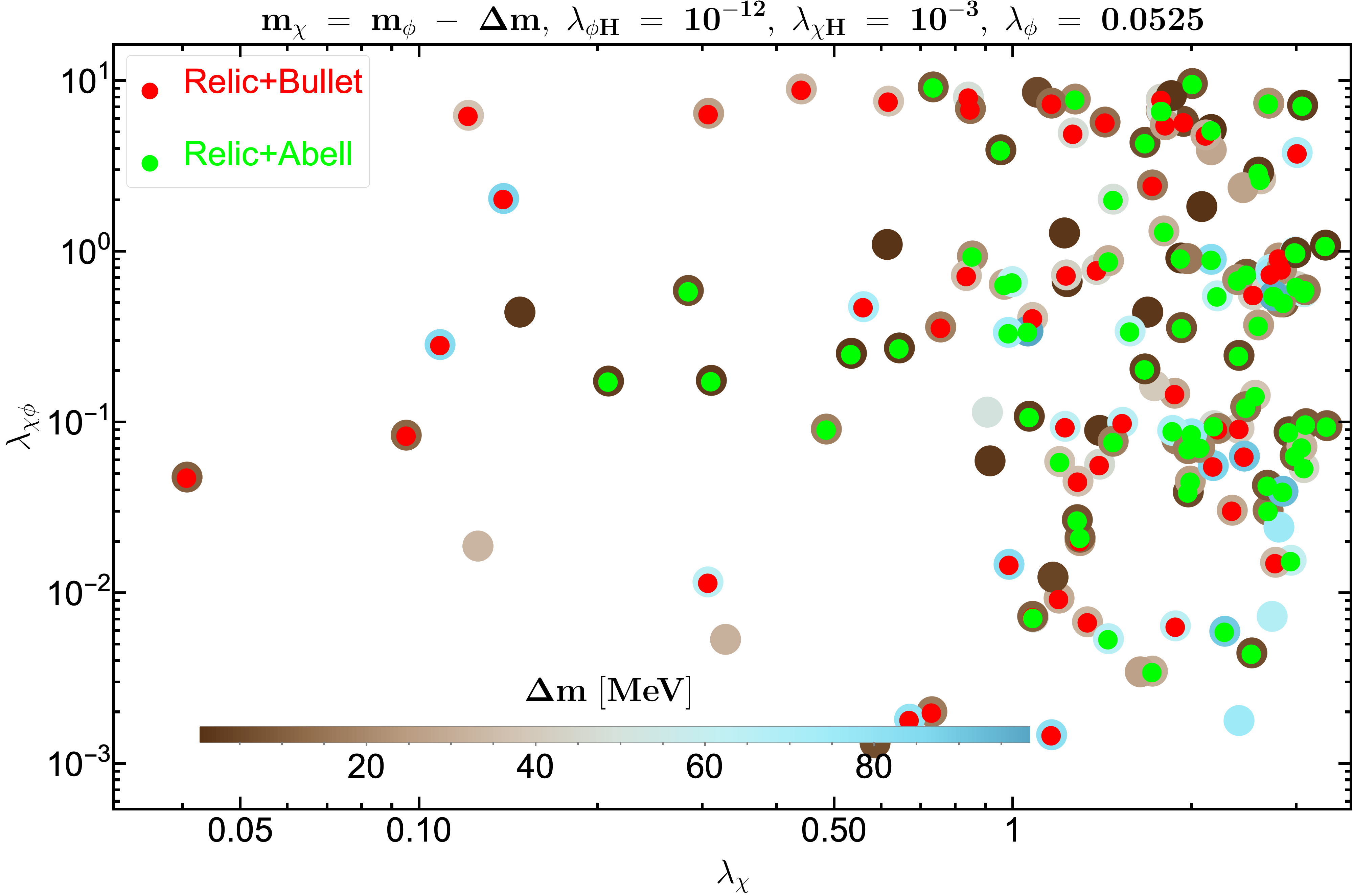}\label{fig:-pfimp-simp_chilphi-lchi-lchiphi-dm}}
\caption{Figs\,.~\ref{fig:-pfimp-simp_chigphi-mchi-lchi-lchiphi}, \ref{fig:-pfimp-simp_chilphi-mchi-lchi-lchiphi} and \ref{fig:-pfimp-simp_chigphi-mchi-mu3-lchiphi}, \ref{fig:-pfimp-simp_chilphi-mchi-mu3-lchiphi} are represent the DM relic allowed parameter space in $m_{\chi}-\lambda_{\chi}$ and $m_{\chi}-\mu_3$ plane, respectively. The variation of portal coupling $\lambda_{\chi\phi}$ is shown by the BrownCyanTones color bar. The red and green points represent the points within the Bullet and Abell cluster self-interaction limit. Figs\,.~\ref{fig:-pfimp-simp_chigphi-lchi-lchiphi-dm}, \ref{fig:-pfimp-simp_chilphi-lchi-lchiphi-dm} represents the relic allowed points in $\lambda_{\chi}-\lambda_{\chi\phi}$ plane and BrownCyanTones color bar shows the DM mass separation.}
\label{fig:scan1}
\end{figure}

We will provide a short account of the results obtained from the parameter space scan here. 
In the two-component SIMP-pFIMP scenario, total relic density $(\rm\Omega^{obs}_{DM})$ is gathered from both SIMP and pFIMP, obtained from 
the solution of cBEQ in eq\,.~\ref{eq:cbeq}. In our model, as described by eq\,.~\ref{eq:model}, free parameters relevant for the cosmological evolution are 
$m_{\chi},~m_{\phi},~\lambda_{\chi},~\mu_3$ and $\lambda_{\chi\phi}$. Other parameters like $\lambda_{\chi H}\sim 10^{-3}$ and $\lambda_{\phi H}\sim 10^{-12}$ 
are kept fixed to maintain the criteria of keeping $\chi$ as SIMP and $\phi$ as pFIMP. Also, $\lambda_{\chi H}$ and $\lambda_{\phi H}$ do not play a direct 
role in the relic density of the DM components. In fig\,.~\ref{fig:scan1}, we have shown a few variations in the parameters space scan. 

DM self interaction plays a crucial role here. In the SIMP-pFIMP scenario, DM self-interaction cross-section over DM mass 
$(\rm \sigma_{self}/m_{DM})$ depends not only on their self-scattering cross-sections but also on 
their individual relic densities, which are restricted by Bullet cluster and Abell cluster observations, see eq\,~\ref{eq:dm-self}. The parameters $\lambda_{\chi H}$ and 
$\lambda_{\phi}$ play a significant role in DM self-interaction processes. If the pFIMP relic is dominant over SIMP, then $\sigma_{\rm self}$ mostly 
depends on the $\sigma_{\phi\phi\to\phi\phi}$, {\it i.e\,.~}$\lambda_{\phi}$, but, in case of dominant SIMP, the self scattering cross-section depends primarily on 
{\it i.e.~}$\lambda_{\chi}$ and $\lambda_{\chi H}$ couplings.

In figs.~\ref{fig:-pfimp-simp_chigphi-mchi-lchi-lchiphi} and ~\ref{fig:-pfimp-simp_chilphi-mchi-lchi-lchiphi}, we have plotted the relic density allowed parameter space in 
$m_{\chi}-\lambda_{\chi}$ plane for both the mass hierarchies. SIMP is heavier in the left plot, while pFIMP is heavier on the right. The main parameter that is varied here 
is the coupling relevant for conversion cross-section, i.e. $\lambda_{\chi\phi}$, as shown in the colour bar. Other parameters kept fixed are mentioned in figure heading, 
they are responsible for keeping the SIMP and pFIMP limit intact and respect unitarity, vacuum stability limits. All the scattered points respect relic density constraint. Those in 
red respect self interaction limit from Bullet cluster, while those in green respect the limit from Abell cluster. When SIMP is heavier 
(fig. \ref{fig:-pfimp-simp_chigphi-mchi-lchi-lchiphi}), pFIMP freezes out before and SIMP relic is smaller and has a subdominant contribution to the 
total relic density due to modified freeze out. In such circumstances, self interaction limit is less constrained for SIMP and allows $\lambda_{\chi}\sim \times 10^{-2}$. For a lighter 
SIMP, the constraint is stricter $\lambda_{\chi}\sim \times 10^{-1}$. $\lambda_{\phi}\sim 5.25\times 10^{-2}$ helps the pFIMP to obey self-interaction bound.

In figs\,.~\ref{fig:-pfimp-simp_chigphi-mchi-mu3-lchiphi} and \ref{fig:-pfimp-simp_chilphi-mchi-mu3-lchiphi}, we have plotted the relic allowed parameter space in 
$m_{\chi}-\mu_{3}$ planes. Red points respect self interaction limit from Bullet cluster, while those in green respect the limit from Abell cluster. Almost all the 
points in the full mass range obeys this limit. We further see a correlation between $m_{\chi}$ and $\mu_{3}$, roughly following $\mu_3 \ge 2m_\chi$. 
As before, $\lambda_{\chi\phi}$ is varied as shown in the colour bar.
For larger $\lambda_{\chi\phi}$, when pFIMP is heavier (right hand plot), it shows modified freeze out, and its relic becomes sub-dominant 
in total DM relic. Under this circumstance, only SIMP self-interaction contributes dominantly to DM self-interaction bound (Note, we kept $\lambda_{\chi H}$ is constant). 
$\lambda_{\chi H}\sim 10^{-3}$ is preferred for the allowed parameter space to obey both Bullet and Abell cluster bound while $\lambda_{\phi}$ is free. Lower mass 
limit $m_\chi \lesssim$ 15 MeV is disfavoured from the self interaction limit. 

In figs\,.~\ref{fig:-pfimp-simp_chigphi-lchi-lchiphi-dm} and \ref{fig:-pfimp-simp_chilphi-lchi-lchiphi-dm}, we have shown the relic density allowed points with 
DM mass variation in the color bar in $\lambda_\chi-\lambda_{\chi\phi}$ plane. As such there is no specific correlation between these parameters to be observed to 
yield correct relic abundance and self interaction limits. We see that the maximum available DM mass separation is up to $5$ MeV for 
$m_{\chi}>m_{\phi}$, but more relaxed (up to $50$ MeV) for the opposite mass hierarchy. This is an important result of this pFIMP-SIMP scenario, to agree with the model-independent analysis of pFIMP in the presence of any thermal DM \cite{Bhattacharya:2022dco, Bhattacharya:2022vxm}.

Before concluding this section, let us briefly summarise the outcome of this analysis. The two-component DM scenario in the presence of SIMP has not been studied much 
excepting \cite{Choi:2021yps, Ho:2022erb}. Here, we have studied pFIMP dynamics in the presence of SIMP that helps a feebly coupled DM to equilibrate to thermal bath 
and freeze out via substantial interaction with SIMP. The major outcomes of this SIMP-pFIMP analysis are that relic density allowed parameter space of this model surpasses the single component SIMP allowed range, in particular, the SIMP mass is allowed up to $\sim 50$ MeV when $m_{\phi}>m_{\chi}$. Second, DM self-interaction provides the most dominant constraint, in particular, $\lambda_{\phi}$ and $\lambda_{\chi H}$ couplings, which are free parameters in context of DM relic density gets constrained by DM self-interaction bound.

\subsection{Detection possibility}
\label{sec:-detection}

The required condition to keep the SIMP in kinematic equilibrium is $\Gamma_{\chi\chi^*\to f\overline{f}}\lesssim\Gamma_{3\chi\to2\chi}\lesssim\Gamma_{\chi f\to \chi f}$, 
which has motivated us to choose $\lambda_{\chi H}$ as free parameter (in single component) as long as it doesn't become too high such that the $\rm DM~DM \to SM~SM$ annihilation dominates over the self-annihilation. As this parameter is suppressed, it doesn't have a significant role in SIMP freeze out and relic density, 
but this might be constrained by presently available SIMP-electron scattering bound like XENON1T \cite{XENON:2019gfn}, CRESST-III \cite{CRESST:2019jnq}, DAMIC-M \cite{DAMIC-M:2023gxo}, DarkSide-50 \cite{DarkSide:2018bpj}, ALETHEIA \cite{Liao:2022thr} and DarkSide-20k \cite{DarkSide-20k:2017zyg}, LDMX \cite{LDMX:2018cma} (proposed) etc.

Unfortunately, in the case of our two-component real and complex scalar pFIMP-SIMP model, it is difficult to detect  both the components via 
direct or indirect detection due to small values of $\lambda_{\chi H}$ with detector's low sensitivity as far as the current and immediate future projection goes. 
The DM-electron cross-section mostly depend upon the mediator mass $m_h$ and Higgs-DM portal coupling $\lambda_{\chi H}$ for SIMP. Apart from small $\lambda_{\chi H}$, 
the presence of a heavy $(\sim\rm GeV)$ mediator decreases the DM-electron scattering cross-section.
As we already know \cite{Bhattacharya:2022dco, Bhattacharya:2022vxm}, the pFIMP detection is possible only via thermal DM loop-mediated interaction with the visible sector, 
so the difficulty in SIMP detection transmits to the difficulty in pFIMP searches as well, although low mass pFIMP takes part in electron scattering. The pFIMP-electron scattering cross-section depends upon the parameters $\lambda_{\chi \phi},~\lambda_{\chi H}$ and the mediator mass ($\rm m_h\sim~125~GeV)$. In the case of $\rm MeV$ DM (SIMP or pFIMP), the DM-electron scattering cross-section significantly suffers from the presence of heavy mediator.

Therefore detection possibility of such a framework emerges under the possibility of a low mediator mass $(\rm \sim MeV)$. For that the model needs to be augmented further. 
 Besides the particle content of two scalar fields contributing to pFIMP-WIMP model, we consider a vector-like lepton (VLL) $\psi$ with weak hypercharge ${\tt Y}=-1$ \cite{Athron:2021iuf, Kawamura:2020qxo}, charged under $\mathbb{Z}_3$ having transformation like $\psi\to\omega\psi$. For simplicity, one may assume that this 
 VLL fermion only couples to the first right-handed lepton generation to avoid the strong bounds from the lepton flavour-violating process. The newly extended Lagrangian 
 can be written as,
\bea
\mathcal{L}_{\psi}=\overline{\psi}\left[i\gamma^{\mu}\left(\partial_{\mu}+ig^{\prime}{\tt Y} B_{\mu}\right)-m_{\psi}\right]\psi\,-\,\left(\mathtt{y} \overline{\psi}e_R\chi+\rm h.c.\right)\,,
\label{eq:psi}
\eea
where $g^{\prime}=(2/v)\sqrt{m_Z^2-m_W^2}$ is the $\rm U(1)_{\tt Y}$ gauge coupling. $\psi$ is a vector-like lepton (VLL) with weak hypercharge ${\tt Y}=-1$ \cite{Athron:2021iuf, Kawamura:2020qxo} and charged under $\mathbb{Z}_3$ where transformation is like $\psi\to\omega\psi$. The perturbative limit on real Yukawa coupling is $\mathtt{y}<\sqrt{4\pi}$ and $m_{\psi}>m_{\chi}+m_e$ to ensure the $\chi$ stability. The most important mass bound on this charged particle mass comes from LEP \cite{L3:2001xsz, ALEPH:2002nwp, OPAL:2003nhx, DELPHI:2003uqw, L3:2003fyi} to be $\gtrsim 103.5$ GeV. In that way, 
this also becomes a heavy mediator. The detection possibility and phenomenology of such a framework are discussed in \cite{Lahiri:2024rxc}, with a focus on dark matter masses in the GeV range.
\section{Summary and conclusions}
\label{sec:5}

Multicomponent DM is a viable option to address observed DM characteristics. Such frameworks can be constituted of any kind of DM particle, like WIMP, FIMP, or SIMP. 
pFIMP is an interesting possibility which exists only in presence of a thermal DM component, which relies on sizable conversion process to equilibrate with thermal bath 
and freeze out. In this article, we have discussed pFIMP possibility in presence of SIMP. This is the first kind of an analysis to address such a pFIMP-SIMP framework.

In order to have a clear picture of what two component framework does, we first discuss the simplest SIMP framework, 
complete analytic solution of SIMP and compare it with the numerical solution using Mathematica to 
show a close comparison. Notably, this analytical solution does better than the previously obtained one \cite{Bhattacharya:2019mmy} to agree with the numerical solution, 
which shows the relic density allowed parameter space lies within $10-200$ MeV of SIMP mass, some portions of which 
are excluded by the presently available DM self-interaction bound from Bullet and Abell clusters. 

To address pFIMP dynamics in presence of SIMP, we first adopt a model independent analysis by solving coupled BEQs, one of which addresses SIMP freeze out, the other 
pFIMP freeze out. For this part, the responsible thermal average cross-sections are taken as numerical numbers. The solution of cBEQ, as shown in fig\,.~\ref{fig:-sol-cbeq}, 
confirms the pFIMP dynamics in the presence of SIMP, when conversion cross-section is varied from feeble to weak strength. We show that a pure FIMP having freeze in can be 
brought to thermal bath and freeze out when the conversion to accompanying SIMP is sizeable. In pFIMP limit, the depletion within the dark sector $\rm 3SIMP\to 2SIMP$ 
might be comparable to the conversion process $\rm 2SIMP\to2pFIMP$ without affecting the SIMP condition. 

In this analysis, we assumed that SM and dark bath are in kinetic equilibrium via elastic scattering of DM with bath particles. This elastic scattering rate measures the momentum exchange rate to/from SM bath from/to the dark sector. If this rate is larger than the Hubble expansion rate, the dark sector and thermal bath are always in kinetic equilibrium, and both sectors follow the same temperature (bath temperature). This has been validated in the appendix for a typical benchmark. Otherwise, the dark sector temperature would be different from the SM bath. It will be an interesting exercise to take up such an analysis in the context of the SIMP-pFIMP setup.

We next discuss the simplest pFIMP-SIMP model with extension of SM with two scalar singlets (one real, one complex) stabilised by $\mathbb{Z}_2\otimes \mathbb{Z}_3$ symmetry. 
We chose the model parameter in such a way that the real scalar behaves as pFIMP ($\phi$), while the complex scalar ($\chi$) acts like SIMP. 
Out of the DM masses, self couplings, portal couplings, and conversion couplings, the conversion coupling $\lambda_{\chi\phi}$ helps the pFIMP reach thermal equilibrium. 
The parameter space is crucially constrained by the self interaction bounds. Interestingly, couplings like $\rm \lambda_{\phi},~and~\lambda_{\chi H}$, which play 
insignificant role in DM freeze out, gets constrained by DM self-interaction. As usual, DM mass splittings get constrained to yield pFIMP solution, but different 
depending on mass hierarchy. For example, $\Delta m \lesssim \rm 2~MeV$ for heavier SIMP, and $\Delta m\lesssim \rm 100~MeV$ when pFIMP is heavier. 
Also, relic density allowed parameter space of this model surpasses the single component SIMP allowed range, in particular, the SIMP mass is allowed 
up to $\sim 50$ MeV when $m_{\phi}>m_{\chi}$.

Detectability of pFIMP-SMIP model is difficult as the portal coupling of both pFIMP and SIMP with SM Higgs is suppressed to address their freeze outs. Secondly, their 
interaction with SM is mediated by SM Higgs, which is way heavier than the DM mass ($\sim$MeV), making the cross-section further subdued. 
We thus propose a minimal extension of the dark sector by the inclusion of a charged vector-like lepton. This is primarily assumed to interact with electrons only, to avoid 
 conflicts with lepton flavor violation constraints. This provides a different mediator to interact with direct or indirect searches. The presence of such new Yukawa interaction terms opens up direct (via electron scattering) and indirect (annihilation into electron and photon pair) search prospects of the pFIMP-SIMP scenario.
\acknowledgments

DP thanks Heptagon, IITG, for valuable discussions. SB and JT thanks MTP project at IIT Guwahati, where the work was initiated. 

\appendix

\section{The semi-analytical solution of a SIMP}
\label{app:A}

The Boltzmann equation for the number density $n_{s}$ of the SIMP type DM $\rm (s)$,
\bea
\dot{n_s}+3 Hn_s=-\left(n_s^3-n_s^2n_{s}^{\rm eq}\right)\langle \sigma v^2\rangle_{3\to2}-\left(n_s^2-n^{\rm eq^2}_{s}\right)\langle \sigma v\rangle_{\rm ann}\,.
\eea
As the necessity condition for SIMP is $\Gamma_{3\to 2}\gtrsim\Gamma_{\rm ann}$ so we can safely neglect the $\langle \sigma v\rangle_{\rm ann}$ term, and the BEQ becomes.
\bea
\dfrac{dY_s}{dx}=-\dfrac{n_f}{x^5}\left(Y_s^3-Y_s^2Y_{s}^{\rm eq}\right)\,,
\label{eq:beq1}
\eea
where $x=m_s/T$, $n_f=\dfrac{ {\bf s}^2x^4}{\mathcal{H}(x)}\langle \sigma v^2\rangle_{3\to2}$ with $s$ and $\mathcal{H}(x)$ are entropy density and Hubble parameter, respectively. The equilibrium yield of SIMP is given by Maxwell-Boltzmann distribution, as during freeze out SIMP is in the non-relativistic regime, $Y_{s}^{\rm eq}=0.145 ({g_s}/{g_{\star}^{\bf s}})x^{3/2}e^{-x}=A ~x^{3/2}e^{-x}$ where $A=0.145 ({g_s}/{g_{\star}^{s}})$. Another important assumption is that the entropy and matter degrees of freedom are nearly equal during SIMP freezes out ($x_{\rm FO}\sim 25$) and constant $g_{\star}^ {\bf s} \simeq g_{\star}^{\rho}\sim 10$.

To solve the BEQ analytically, it would be convenient to divide the whole scenario into three different regions on value of $x$: {\bf Region I:} $x\ll x_{\rm FO}$, {\bf Region II:} $x\simeq x_{\rm FO}$ and {\bf Region III:} $x\gg x_{\rm FO}$. Let us define the difference of SIMP yield from its equilibrium yield as $\Delta=Y_s-Y_{s}^{\rm eq}$ and eq\,~\ref{eq:beq1} becomes,
\bea
\dfrac{dY_{s}^{\rm eq}}{dx}+\dfrac{d \Delta}{dx}=-\dfrac{n_f}{x^5}\Delta\left(\Delta+Y_{s}^{\rm eq}\right)^2\,.
\label{eq:beq2}
\eea
\begin{itemize}
\item{\bf Region I\\}
For $x\ll x_{\rm FO}$, $\dfrac{d \Delta}{dx}$ is negligible and eq\,.~\ref{eq:beq2} becomes,
\bea
\dfrac{dY_{s}^{\rm eq}}{dx}=-\dfrac{n_f}{x^5}\Delta\left(\Delta+Y_{s}^{\rm eq}\right)^2\,.
\label{eq:beq3}
\eea
\item{\bf Region II\\}
In the vicinity of $x\simeq x_{\rm FO}$, we may further assume $\Delta=c ~Y_{s}^{\rm eq}(x_{\rm FO})$ where $c$ is unknown constants whose values are determined by matching the analytical result with the numerical one. Substituting in eq\,.~\ref{eq:beq3}, we get
\bea
\dfrac{dY_{s}^{\rm eq}}{dx}\bigg|_{x=x_{\rm FO}}=-\dfrac{n_f}{x_{\rm FO}^5}c\left(c+1\right)^2Y_{s}^{\rm eq^3}(x_{\rm FO})\,.
\label{eq:beq4}
\eea
After simplification, we get,
\bea
x_{\rm FO}^{2}e^{2x_{\rm FO}}-\dfrac{3}{2}x_{\rm FO}e^{2x_{\rm FO}}=\underbrace{n_f ~A^2~c\left(c+1\right)^2}_{P}\,.
\label{eq:beq5}
\eea
The solution of eq\,.~\ref{eq:beq5} gives us the freeze out point of SIMP is,
\bea
x_{\rm FO}\approx\ln \sqrt{P}-\ln\ln\sqrt{P}\,.
\eea

\item{\bf Region III\\}
When $x\gg x_{\rm FO}$, then, $Y_{s}^{\rm eq}$ and $\dfrac{dY_{s}^{\rm eq}}{dx}$ are exponentially suppressed, so, $\Delta \gg Y_{s}^{\rm eq}$ and eq\,.~\ref{eq:beq2} becomes,
\bea
\dfrac{d \Delta}{dx}=-\dfrac{n_f}{x^5}\Delta^3\,.
\label{eq:beq6}
\eea
After solving the differential equation between $x_{\rm FO}$ to $x$ with $\Delta(x)\gg \Delta(x_{\rm FO})$, we get,
\bea
\Delta(x)=\left[\dfrac{1}{ \Delta(x_{\rm FO})^2}-\dfrac{n_f}{2}\left(\dfrac{1}{x^4}-\dfrac{1}{x^4_{\rm FO}}\right) \right]^{-\dfrac{1}{2}}\,.
\label{eq:beq7}
\eea
Following $\Delta(x_{\rm FO})\approx c Y_{s}^{\rm eq}(x_{\rm FO})$ at $x=x_{\rm CMB}=m_s/T_{\rm CMB}$, the SIMP yields is
\bea
Y_s(x_{\rm CMB})\approx \Delta(x_{\rm CMB})\approx \left[\dfrac{1}{ \Delta(x_{\rm FO})^2}-\dfrac{n_f}{2}\left(\dfrac{1}{x_{\rm CMB}^4}-\dfrac{1}{x^4_{\rm FO}}\right) \right]^{-\dfrac{1}{2}}\,.
\label{eq:beq7}
\eea
\end{itemize}

\begin{figure}[htb!]
\centering
\subfloat[]{\includegraphics[width=0.475\linewidth]{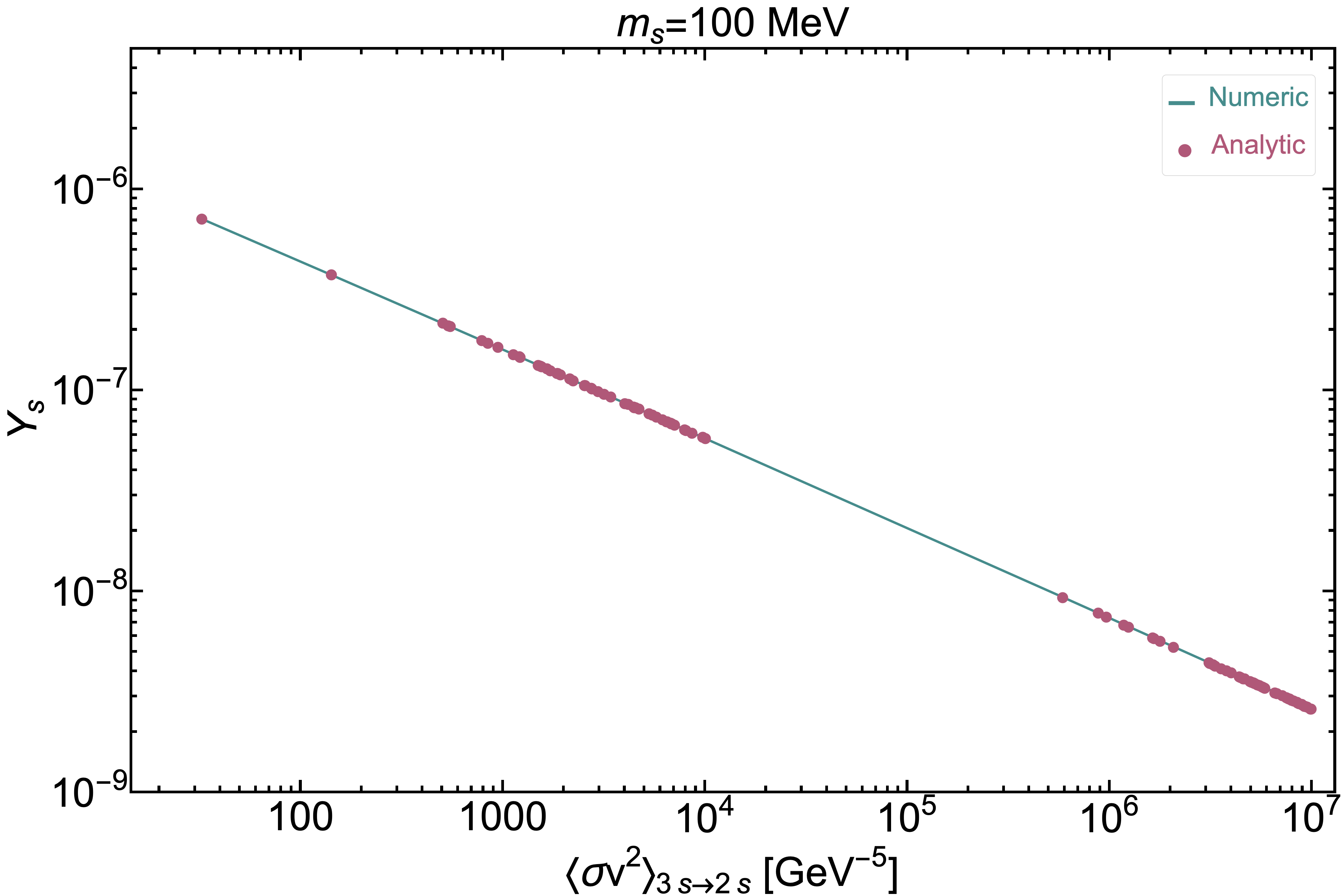}\label{fig:-simp_analytic_sigma}}\quad
\subfloat[]{\includegraphics[width=0.475\linewidth]{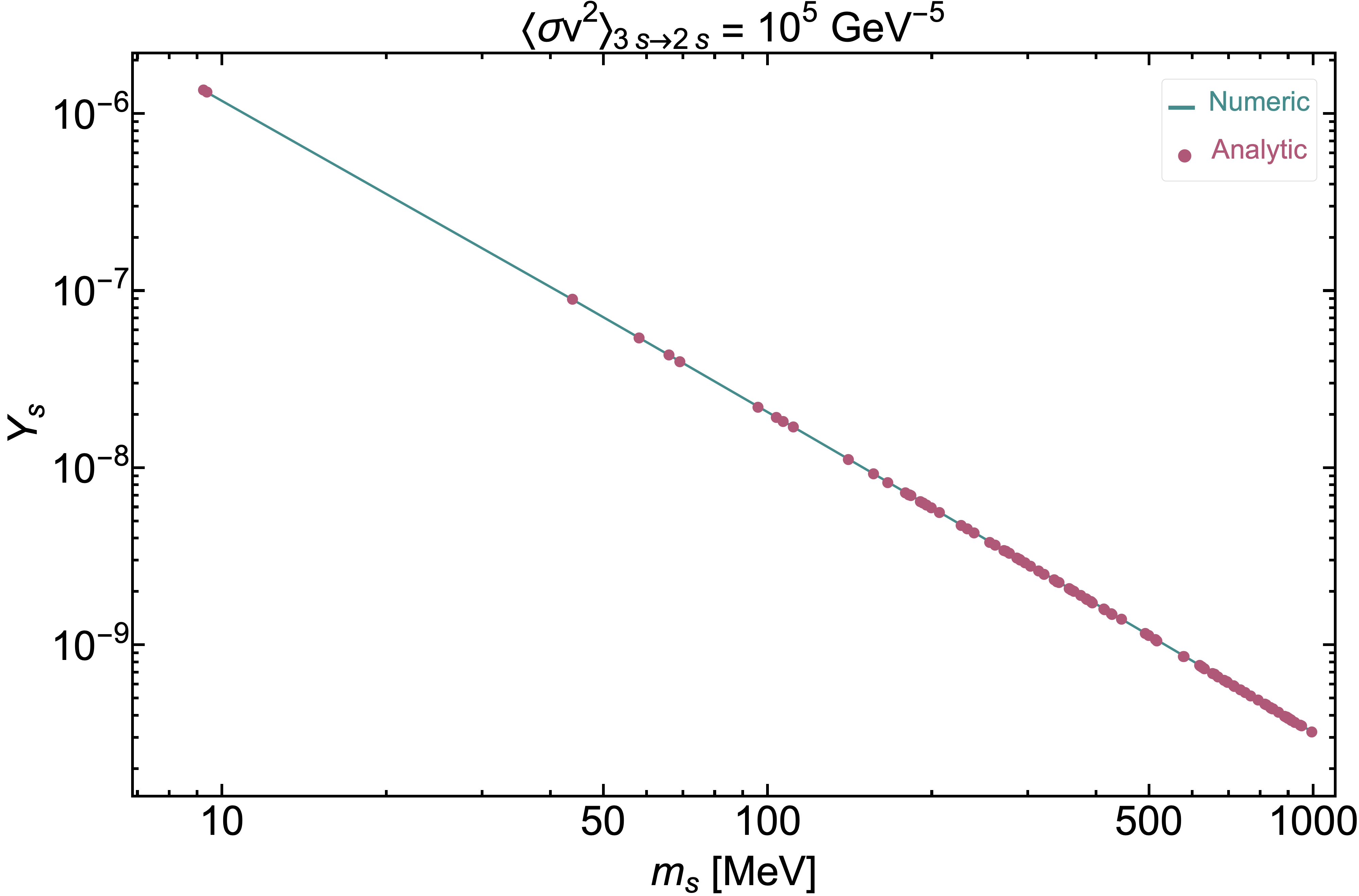}\label{fig:-simp_analytic_ms}}
\caption{A comparison plot analytic vs numeric solution of BEQ where we have considered $g_{\star}^{\bf s} =g_{\star}^{\rho}=10.75$, $g_s=2$ (for complex scalar) and have chosen $c(c+1)^2=4.5$.}\label{fig:sol_ind}
\end{figure}

\section{Possible Feynman diagrams related to DM phenomenology}
\label{app:B}
We have calculated the squared matrix amplitude for non-relativistic DMs using FeynCalc \cite{Shtabovenko:2023idz} and CalcHEP \cite{Belyaev:2012qa}.
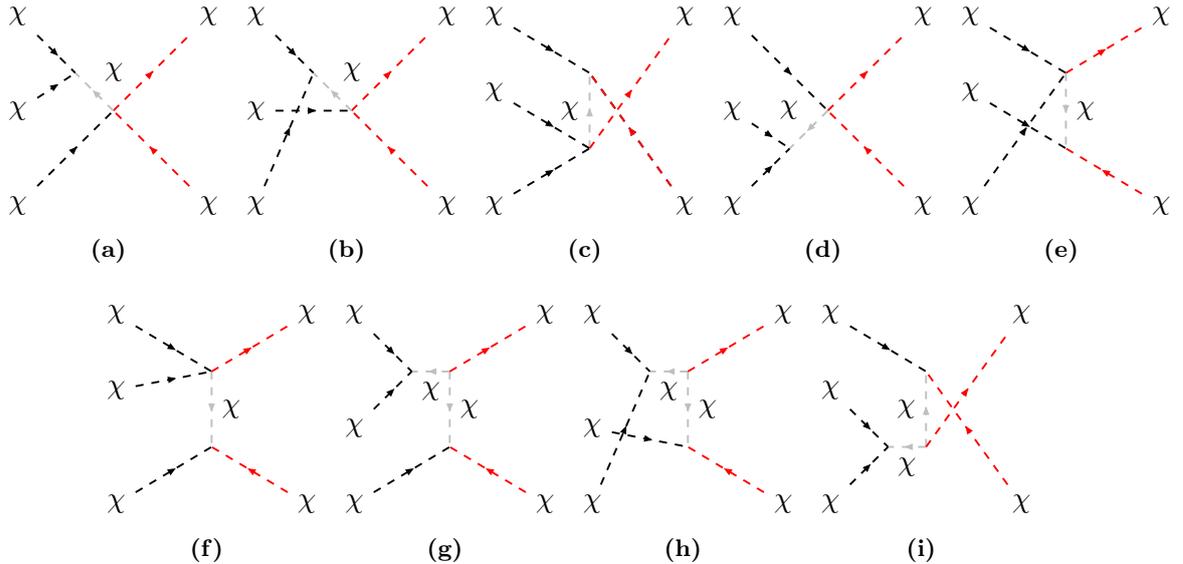
\begin{figure}[htb!]
\centering
\subfloat[]{
\begin{tikzpicture}
\begin{feynman}
\vertex(a);
\vertex[above left=0.5cm and 0.5cm of a] (a3);
\vertex[ left=1cm of a] (a4){\(\chi\)};
\vertex[above left=1cm and 1cm of a] (a1){\(\chi\)};
\vertex[below left=1cm and 1cm of a] (a2){\(\chi\)};
\vertex[above right=1cm and 1cm of a] (b1){\(\chi\)};
\vertex[below right=1cm and 1cm of a] (b2){\(\chi\)};
\diagram*{
(a2) -- [ line width=0.25mm,charged scalar, arrow size=0.7pt,edge label={\(\rm \)},style=black] (a),
(a) -- [ line width=0.25mm,charged scalar, arrow size=0.7pt,edge label={\(\rm \)},style=red] (b1),
(b2) -- [ line width=0.25mm,charged scalar, arrow size=0.7pt,edge label={\(\rm \)},style=red] (a),
(a1) -- [ line width=0.25mm,charged scalar, arrow size=0.7pt,edge label={\(\rm \)},style=black] (a3),
(a4) -- [ line width=0.25mm,charged scalar, arrow size=0.7pt,edge label={\(\rm \)},style=black] (a3),
(a) -- [ line width=0.25mm,charged scalar, arrow size=0.7pt,style=gray!50,edge label'={\(\rm\textcolor{black}{ \chi}\)}] (a3)};
\end{feynman}
\end{tikzpicture}\label{fig:1a}}
\subfloat[]{
\begin{tikzpicture}
\begin{feynman}
\vertex(a);
\vertex[above left=0.5cm and 0.5cm of a] (a3);
\vertex[ left=1cm of a] (a4){\(\chi\)};
\vertex[above left=1cm and 1cm of a] (a1){\(\chi\)};
\vertex[below left=1cm and 1cm of a] (a2){\(\chi\)};
\vertex[above right=1cm and 1cm of a] (b1){\(\chi\)};
\vertex[below right=1cm and 1cm of a] (b2){\(\chi\)};
\diagram*{
(a2) -- [ line width=0.25mm,charged scalar, arrow size=0.7pt,edge label={\(\rm \)},style=black] (a3),
(a) -- [ line width=0.25mm,charged scalar, arrow size=0.7pt,edge label={\(\rm \)},style=red] (b1),
(b2) -- [ line width=0.25mm,charged scalar, arrow size=0.7pt,edge label={\(\rm \)},style=red] (a),
(a1) -- [ line width=0.25mm,charged scalar, arrow size=0.7pt,edge label={\(\rm \)},style=black] (a3),
(a4) -- [ line width=0.25mm,charged scalar, arrow size=0.7pt,edge label={\(\rm \)},style=black] (a),
(a) -- [ line width=0.25mm,charged scalar, arrow size=0.7pt,style=gray!50,edge label'={\(\rm\textcolor{black}{ \chi}\)}] (a3)};
\end{feynman}
\end{tikzpicture}\label{fig:1b}}
\subfloat[]{
\begin{tikzpicture}
\begin{feynman}
\vertex(a);
\vertex[below=1cm of a] (b);
\vertex[above left=0.5cm and 1cm of a] (a2){\(\chi\)};
\vertex[above right=0.5cm and 1cm of a] (b1){\(\chi\)};
\vertex[below right=0.5cm and 1cm of b] (a1){\(\chi\)};
\vertex[above left=0.5cm and 1cm of b] (b2){\(\chi\)};
\vertex[below left=0.5cm and 1cm of b] (b3){\(\chi\)};
\diagram*{
(b) -- [ line width=0.25mm,charged scalar, arrow size=0.7pt,style=gray!50,edge label={\(\rm\textcolor{black}{ \chi}\)}] (a),
(a2) -- [ line width=0.25mm,charged scalar, arrow size=0.7pt,edge label={\(\rm \)},style=black] (a),
(a1) -- [ line width=0.25mm,charged scalar, arrow size=0.7pt,edge label={\(\rm \)},style=black] (a),
(b2) -- [ line width=0.25mm,charged scalar, arrow size=0.7pt,edge label={\(\rm \)},style=black] (b),
(b3) -- [ line width=0.25mm,charged scalar, arrow size=0.7pt,edge label={\(\rm \)},style=black] (b),
(a1) -- [ line width=0.25mm,charged scalar, arrow size=0.7pt,edge label={\(\rm \)},style=red] (a),
(b) -- [ line width=0.25mm,charged scalar, arrow size=0.7pt,edge label={\(\rm \)},style=red] (b1)};
\end{feynman}
\end{tikzpicture}\label{fig:1c}}
\subfloat[]{
\begin{tikzpicture}
\begin{feynman}
\vertex(a);
\vertex[below left=0.5cm and 0.5cm of a] (a3);
\vertex[left=1cm of a] (a4){\(\chi\)};
\vertex[above left=1cm and 1cm of a] (a1){\(\chi\)};
\vertex[below left=1cm and 1cm of a] (a2){\(\chi\)};
\vertex[above right=1cm and 1cm of a] (b1){\(\chi\)};
\vertex[below right=1cm and 1cm of a] (b2){\(\chi\)};
\diagram*{
(a2) -- [ line width=0.25mm,charged scalar, arrow size=0.7pt,edge label={\(\rm \)},style=black] (a3),
(a) -- [ line width=0.25mm,charged scalar, arrow size=0.7pt,edge label={\(\rm \)},style=red] (b1),
(b2) -- [ line width=0.25mm,charged scalar, arrow size=0.7pt,edge label={\(\rm \)},style=red] (a),
(a1) -- [ line width=0.25mm,charged scalar, arrow size=0.7pt,edge label={\(\rm \)},style=black] (a),
(a4) -- [ line width=0.25mm,charged scalar, arrow size=0.7pt,edge label={\(\rm \)},style=black] (a3),
(a) -- [ line width=0.25mm,charged scalar, arrow size=0.7pt,style=gray!50,edge label'={\(\rm\textcolor{black}{ \chi}\)}] (a3)};
\end{feynman}
\end{tikzpicture}\label{fig:1d}}
\subfloat[]{
\begin{tikzpicture}
\begin{feynman}
\vertex(a);
\vertex[ below=1cm of a] (b);
\vertex[above left=0.5cm and 1cm of a] (a2){\(\chi\)};
\vertex[above right=0.5cm and 1cm of a] (b1){\(\chi\)};
\vertex[below right=0.5cm and 1cm of b] (a1){\(\chi\)};
\vertex[above left=0.5cm and 1cm of b] (b2){\(\chi\)};
\vertex[below left=0.5cm and 1cm of b] (b3){\(\chi\)};
\diagram*{
(a) -- [ line width=0.25mm,charged scalar, arrow size=0.7pt,style=gray!50,edge label={\(\rm\textcolor{black}{ \chi}\)}] (b),
(a2) -- [ line width=0.25mm,charged scalar, arrow size=0.7pt,edge label={\(\rm \)},style=black] (a),
(b2) -- [ line width=0.25mm,charged scalar, arrow size=0.7pt,edge label={\(\rm \)},style=black] (b),
(b3) -- [ line width=0.25mm,charged scalar, arrow size=0.7pt,edge label={\(\rm \)},style=black] (a),
(a) -- [ line width=0.25mm,charged scalar, arrow size=0.7pt,edge label={\(\rm \)},style=red] (b1),
(a1) -- [ line width=0.25mm,charged scalar, arrow size=0.7pt,edge label={\(\rm \)},style=red] (b)};
\end{feynman}
\end{tikzpicture}\label{fig:1e}}

\subfloat[]{
\begin{tikzpicture}
\begin{feynman}
\vertex(a);
\vertex[ below=1cm of a] (b);
\vertex[above left=0.5cm and 1cm of a] (a2){\(\chi\)};
\vertex[above right=0.5cm and 1cm of a] (b1){\(\chi\)};
\vertex[below right=0.5cm and 1cm of b] (a1){\(\chi\)};
\vertex[above left=0.5cm and 1cm of b] (b2){\(\chi\)};
\vertex[below left=0.5cm and 1cm of b] (b3){\(\chi\)};
\diagram*{
(a) -- [ line width=0.25mm,charged scalar, arrow size=0.7pt,style=gray!50,edge label={\(\rm\textcolor{black}{ \chi}\)}] (b),
(a2) -- [ line width=0.25mm,charged scalar, arrow size=0.7pt,edge label={\(\rm \)},style=black] (a),
(b2) -- [ line width=0.25mm,charged scalar, arrow size=0.7pt,edge label={\(\rm \)},style=black] (a),
(b3) -- [ line width=0.25mm,charged scalar, arrow size=0.7pt,edge label={\(\rm \)},style=black] (b),
(a) -- [ line width=0.25mm,charged scalar, arrow size=0.7pt,edge label={\(\rm \)},style=red] (b1),
(a1) -- [ line width=0.25mm,charged scalar, arrow size=0.7pt,edge label={\(\rm \)},style=red] (b)};
\end{feynman}
\end{tikzpicture}\label{fig:1f}}
\subfloat[]{
\begin{tikzpicture}
\begin{feynman}
\vertex(a);
\vertex[ below=1cm of a] (b);
\vertex[ left=0.5cm of a] (c);
\vertex[above left=0.5cm and 0.5cm of c] (c1){\(\chi\)};
\vertex[below left=0.5cm and 0.5cm of c] (c2){\(\chi\)};
\vertex[above right=0.5cm and 1cm of a] (a2){\(\chi\)};
\vertex[below left=0.5cm and 1cm of b] (b1){\(\chi\)};
\vertex[below right=0.5cm and 1cm of b] (b2){\(\chi\)};
\diagram*{
(a) -- [ line width=0.25mm,charged scalar, arrow size=0.7pt,style=gray!50,edge label={\(\rm\textcolor{black}{ \chi}\)}] (b),
(a) -- [ line width=0.25mm,charged scalar, arrow size=0.7pt,style=gray!50,edge label={\(\rm\textcolor{black}{ \chi}\)}] (c),
(c1) -- [ line width=0.25mm,charged scalar, arrow size=0.7pt,edge label={\(\rm \)},style=black] (c),
(c2) -- [ line width=0.25mm,charged scalar, arrow size=0.7pt,edge label={\(\rm \)},style=black] (c),
(b1) -- [ line width=0.25mm,charged scalar, arrow size=0.7pt,edge label={\(\rm \)},style=black] (b),
(b2) -- [ line width=0.25mm,charged scalar, arrow size=0.7pt,edge label={\(\rm \)},style=red] (b),
(a) -- [ line width=0.25mm,charged scalar, arrow size=0.7pt,edge label={\(\rm \)},style=red] (a2)};
\end{feynman}
\end{tikzpicture}\label{fig:1g}}
\subfloat[]{
\begin{tikzpicture}
\begin{feynman}
\vertex(a);
\vertex[below=1cm of a] (b);
\vertex[left=0.5cm of a] (c);
\vertex[above left=0.5cm and 0.5cm of c] (c1){\(\chi\)};
\vertex[below left=0.5cm and 0.5cm of c] (c2){\(\chi\)};
\vertex[above right=0.5cm and 1cm of a] (a2){\(\chi\)};
\vertex[below left=0.5cm and 1cm of b] (b1){\(\chi\)};
\vertex[below right=0.5cm and 1cm of b] (b2){\(\chi\)};
\diagram*{
(a) -- [ line width=0.25mm,charged scalar, arrow size=0.7pt,style=gray!50,edge label={\(\rm\textcolor{black}{ \chi}\)}] (b),
(a) -- [ line width=0.25mm,charged scalar, arrow size=0.7pt,style=gray!50,edge label={\(\rm\textcolor{black}{ \chi}\)}] (c),
(c1) -- [ line width=0.25mm,charged scalar, arrow size=0.7pt,edge label={\(\rm \)},style=black] (c),
(c2) -- [ line width=0.25mm,charged scalar, arrow size=0.7pt,edge label={\(\rm \)},style=black] (b),
(b1) -- [ line width=0.25mm,charged scalar, arrow size=0.7pt,edge label={\(\rm \)},style=black] (c),
(b2) -- [ line width=0.25mm,charged scalar, arrow size=0.7pt,edge label={\(\rm \)},style=red] (b),
(a) -- [ line width=0.25mm,charged scalar, arrow size=0.7pt,edge label={\(\rm \)},style=red] (a2)};
\end{feynman}
\end{tikzpicture}\label{fig:1h}}
\subfloat[]{
\begin{tikzpicture}
\begin{feynman}
\vertex(a);
\vertex[ above=1cm of a] (b);
\vertex[ left=0.5cm of a] (c);
\vertex[below left=0.5cm and 0.5cm of c] (c1){\(\chi\)};
\vertex[above left=0.5cm and 0.5cm of c] (c2){\(\chi\)};
\vertex[below right=0.5cm and 1cm of a] (a2){\(\chi\)};
\vertex[above left=0.5cm and 1cm of b] (b1){\(\chi\)};
\vertex[above right=0.5cm and 1cm of b] (b2){\(\chi\)};
\diagram*{
(a) -- [ line width=0.25mm,charged scalar, arrow size=0.7pt,style=gray!50,edge label={\(\rm\textcolor{black}{ \chi}\)}] (b),
(a) -- [ line width=0.25mm,charged scalar, arrow size=0.7pt,style=gray!50,edge label={\(\rm\textcolor{black}{ \chi}\)}] (c),
(c1) -- [ line width=0.25mm,charged scalar, arrow size=0.7pt,edge label={\(\rm \)},style=black] (c),
(c2) -- [ line width=0.25mm,charged scalar, arrow size=0.7pt,edge label={\(\rm \)},style=black] (c),
(b1) -- [ line width=0.25mm,charged scalar, arrow size=0.7pt,edge label={\(\rm \)},style=black] (b),
(a2) -- [ line width=0.25mm,charged scalar, arrow size=0.7pt,edge label={\(\rm \)},style=red] (b),
(a) -- [ line width=0.25mm,charged scalar, arrow size=0.7pt,edge label={\(\rm \)},style=red] (b2)};
\end{feynman}
\end{tikzpicture}\label{fig:1i}}
\caption{Self annihilation via $ \chi(p_1) \chi(p_2) \chi(p_3)\rightarrow \chi(p_4) \chi^*(p_5)$ process.}
\label{fig:1}
\end{figure}

\begin{gather}
\overline{|\mathcal{M}_{\ref{fig:1}}|^2}=\dfrac{9\mu_3^2}{4m_{\chi}^8}\left(4\lambda_{\chi} m_{\chi}^2+9\mu_3^2 \right)^2\,.
\end{gather}
and
\begin{gather}
\langle\sigma v^2\rangle_{\chi\chi\chi\to\chi\chi^*}=\dfrac{1}{64\pi m_{\chi}^3}\left(\dfrac{K_1(m_{\chi}/T)}{K_2(m_{\chi}/T)}\right)^3\dfrac{\sqrt{5}}{6}\overline{|\mathcal{M}_{\chi\chi\chi\to\chi\chi^*}|^2}\,.
\end{gather}

\begin{figure}[htb!]
\centering
\subfloat[]{
\begin{tikzpicture}
\begin{feynman}
\vertex(a);
\vertex[below left=0.5cm and 0.5cm of a] (a3);
\vertex[ left=1cm of a] (a4){\(\chi\)};
\vertex[above left=1cm and 1cm of a] (a1){\(\chi\)};
\vertex[below left=1cm and 1cm of a] (a2){\(\chi\)};
\vertex[above right=1cm and 1cm of a] (b1){\(\chi\)};
\vertex[below right=1cm and 1cm of a] (b2){\(\chi\)};
\diagram*{
(a3) -- [ line width=0.25mm,charged scalar, arrow size=0.7pt,edge label={\(\rm \)},style=black] (a2),
(a) -- [ line width=0.25mm,charged scalar, arrow size=0.7pt,edge label={\(\rm \)},style=red] (b1),
(a) -- [ line width=0.25mm,charged scalar, arrow size=0.7pt,edge label={\(\rm \)},style=red] (b2),
(a1) -- [ line width=0.25mm,charged scalar, arrow size=0.7pt,edge label={\(\rm \)},style=black] (a),
(a3) -- [ line width=0.25mm,charged scalar, arrow size=0.7pt,edge label={\(\rm \)},style=black] (a4),
(a3) -- [ line width=0.25mm,charged scalar, arrow size=0.7pt,style=gray!50,edge label'={\(\rm\textcolor{black}{ \chi}\)}] (a)};
\end{feynman}
\end{tikzpicture}\label{fig:2a}}
\subfloat[]{
\begin{tikzpicture}
\begin{feynman}
\vertex(a);
\vertex[below=1cm of a] (b);
\vertex[above left=0.5cm and 1cm of a] (a2){\(\chi\)};
\vertex[below left=0.5cm and 1cm of a] (a3){\(\chi\)};
\vertex[below left=0.5cm and 1cm of b] (b2){\(\chi\)};
\vertex[above right=0.5cm and 1cm of a] (a1){\(\chi\)};
\vertex[below right=0.5cm and 1cm of b] (b1){\(\chi\)};
\diagram*{
(a) -- [ line width=0.25mm,charged scalar, arrow size=0.7pt,style=gray!50,edge label={\(\rm\textcolor{black}{ \chi}\)}] (b),
(a) -- [ line width=0.25mm,charged scalar, arrow size=0.7pt,edge label={\(\rm \)},style=black] (a3),
(a2) -- [ line width=0.25mm,charged scalar, arrow size=0.7pt,edge label={\(\rm \)},style=black] (b),
(b) -- [ line width=0.25mm,charged scalar, arrow size=0.7pt,edge label={\(\rm \)},style=black] (b2),
(b) -- [ line width=0.25mm,charged scalar, arrow size=0.7pt,edge label={\(\rm \)},style=red] (b1),
(a) -- [ line width=0.25mm,charged scalar, arrow size=0.7pt,edge label={\(\rm \)},style=red] (a1)};
\end{feynman}
\end{tikzpicture}\label{fig:2b}}
\subfloat[]{
\begin{tikzpicture}
\begin{feynman}
\vertex(a);
\vertex[below=1cm of a] (b);
\vertex[above left=0.5cm and 1cm of a] (a2){\(\chi\)};
\vertex[below left=0.5cm and 1cm of a] (a3){\(\chi\)};
\vertex[below left=0.5cm and 1cm of b] (b2){\(\chi\)};
\vertex[above right=0.5cm and 1cm of a] (a1){\(\chi\)};
\vertex[below right=0.5cm and 1cm of b] (b1){\(\chi\)};
\diagram*{
(b) -- [ line width=0.25mm,charged scalar, arrow size=0.7pt,style=gray!50,edge label'={\(\rm\textcolor{black}{ \chi}\)}] (a),
(b) -- [ line width=0.25mm,charged scalar, arrow size=0.7pt,edge label={\(\rm \)},style=black] (a3),
(a2) -- [ line width=0.25mm,charged scalar, arrow size=0.7pt,edge label={\(\rm \)},style=black] (a),
(a) -- [ line width=0.25mm,charged scalar, arrow size=0.7pt,edge label={\(\rm \)},style=red] (a1),
(b) -- [ line width=0.25mm,charged scalar, arrow size=0.7pt,edge label={\(\rm \)},style=red] (b1),
(a) -- [ line width=0.25mm,charged scalar, arrow size=0.7pt,edge label={\(\rm \)},style=black] (b2)};
\end{feynman}
\end{tikzpicture}\label{fig:2c}}
\subfloat[]{
\begin{tikzpicture}
\begin{feynman}
\vertex(a);
\vertex[below=1cm of a] (b);
\vertex[above left=0.5cm and 1cm of a] (a2){\(\chi\)};
\vertex[below left=0.5cm and 1cm of a] (a3){\(\chi\)};
\vertex[below left=0.5cm and 1cm of b] (b2){\(\chi\)};
\vertex[above right=0.5cm and 1cm of a] (a1){\(\chi\)};
\vertex[below right=0.5cm and 1cm of b] (b1){\(\chi\)};
\diagram*{
(b) -- [ line width=0.25mm,charged scalar, arrow size=0.7pt,style=gray!50,edge label={\(\rm\textcolor{black}{ \chi}\)}] (a),
(a) -- [ line width=0.25mm,charged scalar, arrow size=0.7pt,edge label={\(\rm \)},style=black] (a3),
(a2) -- [ line width=0.25mm,charged scalar, arrow size=0.7pt,edge label={\(\rm \)},style=black] (a),
(b) -- [ line width=0.25mm,charged scalar, arrow size=0.7pt,edge label={\(\rm \)},style=black] (b2),
(b) -- [ line width=0.25mm,charged scalar, arrow size=0.7pt,edge label={\(\rm \)},style=red] (a1),
(a) -- [ line width=0.25mm,charged scalar, arrow size=0.7pt,edge label={\(\rm \)},style=red] (b1)};
\end{feynman}
\end{tikzpicture}\label{fig:2d}}
\subfloat[]{
\begin{tikzpicture}
\begin{feynman}
\vertex(a);
\vertex[below=1cm of a] (b);
\vertex[above left=0.5cm and 1cm of a] (a2){\(\chi\)};
\vertex[below left=0.5cm and 1cm of a] (a3){\(\chi\)};
\vertex[below left=0.5cm and 1cm of b] (b2){\(\chi\)};
\vertex[above right=0.5cm and 1cm of a] (a1){\(\chi\)};
\vertex[below right=0.5cm and 1cm of b] (b1){\(\chi\)};
\diagram*{
(b) -- [ line width=0.25mm,charged scalar, arrow size=0.7pt,style=gray!50,edge label={\(\rm\textcolor{black}{ \chi}\)}] (a),
(a) -- [ line width=0.25mm,charged scalar, arrow size=0.7pt,edge label={\(\rm \)},style=black] (a3),
(a2) -- [ line width=0.25mm,charged scalar, arrow size=0.7pt,edge label={\(\rm \)},style=black] (a),
(b) -- [ line width=0.25mm,charged scalar, arrow size=0.7pt,edge label={\(\rm \)},style=black] (b2),
(b) -- [ line width=0.25mm,charged scalar, arrow size=0.7pt,edge label={\(\rm \)},style=red] (b1),
(a) -- [ line width=0.25mm,charged scalar, arrow size=0.7pt,edge label={\(\rm \)},style=red] (a1)};
\end{feynman}
\end{tikzpicture}\label{fig:2e}}

\subfloat[]{
\begin{tikzpicture}
\begin{feynman}
\vertex(a);
\vertex[right=1cm of a] (b);
\vertex[left=1cm of a] (a2){\(\chi\)};
\vertex[below left=1cm and 1cm of a] (a3){\(\chi\)};
\vertex[above left=1cm and 1cm of a] (a1){\(\chi\)};
\vertex[above right=1cm and 1cm of b] (b1){\(\chi\)};
\vertex[below right=1cm and 1cm of b] (b2){\(\chi\)};
\diagram*{
(b) -- [ line width=0.25mm,charged scalar, arrow size=0.7pt,style=gray!50,edge label={\(\rm\textcolor{black}{ \chi}\)}] (a),
(a) -- [ line width=0.25mm,charged scalar, arrow size=0.7pt,edge label={\(\rm \)},style=black] (a3),
(a1) -- [ line width=0.25mm,charged scalar, arrow size=0.7pt,edge label={\(\rm \)},style=black] (a),
(a) -- [ line width=0.25mm,charged scalar, arrow size=0.7pt,edge label={\(\rm \)},style=black] (a2),
(a) -- [ line width=0.25mm,charged scalar, arrow size=0.7pt,edge label={\(\rm \)},style=black] (a3),
(b) -- [ line width=0.25mm,charged scalar, arrow size=0.7pt,edge label={\(\rm \)},style=red] (b1),
(b) -- [ line width=0.25mm,charged scalar, arrow size=0.7pt,edge label={\(\rm \)},style=red] (b2)};
\end{feynman}
\end{tikzpicture}\label{fig:2f}}
\subfloat[]{
\begin{tikzpicture}
\begin{feynman}
\vertex(a);
\vertex[right=1cm of a] (b);
\vertex[ left=1cm of a] (a2){\(\chi\)};
\vertex[below left=1cm and 1cm of a] (a3){\(\chi\)};
\vertex[below left=0.5cm and 0.5cm of a] (a4);
\vertex[above left=1cm and 1cm of a] (a1){\(\chi\)};
\vertex[above right=1cm and 1cm of b] (b1){\(\chi\)};
\vertex[below right=1cm and 1cm of b] (b2){\(\chi\)};
\diagram*{
(b) -- [ line width=0.25mm,charged scalar, arrow size=0.7pt,style=gray!50,edge label={\(\rm\textcolor{black}{ \chi}\)}] (a),
(a4) -- [ line width=0.25mm,charged scalar, arrow size=0.7pt,edge label={\(\rm \)},style=gray!50,edge label={\(\rm\textcolor{black}{ \chi}\)}] (a),
(a1) -- [ line width=0.25mm,charged scalar, arrow size=0.7pt,edge label={\(\rm \)},style=black] (a),
(a4) -- [ line width=0.25mm,charged scalar, arrow size=0.7pt,edge label={\(\rm \)},style=black] (a2),
(a4) -- [ line width=0.25mm,charged scalar, arrow size=0.7pt,edge label={\(\rm \)},style=black] (a3),
(b) -- [ line width=0.25mm,charged scalar, arrow size=0.7pt,edge label={\(\rm \)},style=red] (b1),
(b) -- [ line width=0.25mm,charged scalar, arrow size=0.7pt,edge label={\(\rm \)},style=red] (b2)};
\end{feynman}
\end{tikzpicture}\label{fig:2g}}
\subfloat[]{
\begin{tikzpicture}
\begin{feynman}
\vertex(a);
\vertex[below=0.5cm of a] (b);
\vertex[below=0.5cm of b] (c);
\vertex[below left=0.5cm and 1cm of a] (a3){\(\chi\)};
\vertex[above left=0.5cm and 1cm of a] (a1) {\(\chi\)};
\vertex[above right=0.5cm and 1cm of a] (a2){\(\chi\)};
\vertex[below right=0.5cm and 1cm of c] (c2){\(\chi\)};
\vertex[below left=0.5cm and 1cm of c] (c1){\(\chi\)};
\diagram*{
(a) -- [ line width=0.25mm,charged scalar, arrow size=0.7pt,style=gray!50,edge label={\(\rm\textcolor{black}{ \chi}\)}] (b),
(c) -- [ line width=0.25mm,charged scalar, arrow size=0.7pt,edge label={\(\rm \)},style=gray!50,edge label'={\(\rm\textcolor{black}{ \chi}\)}] (b),
(c) -- [ line width=0.25mm,charged scalar, arrow size=0.7pt,edge label={\(\rm \)},style=black] (c1),
(a1) -- [ line width=0.25mm,charged scalar, arrow size=0.7pt,edge label={\(\rm \)},style=black] (b),
(a) -- [ line width=0.25mm,charged scalar, arrow size=0.7pt,edge label={\(\rm \)},style=black] (a3),
(a) -- [ line width=0.25mm,charged scalar, arrow size=0.7pt,edge label={\(\rm \)},style=red] (a2),
(c) -- [ line width=0.25mm,charged scalar, arrow size=0.7pt,edge label={\(\rm \)},style=red] (c2)};
\end{feynman}
\end{tikzpicture}\label{fig:2h}}
\subfloat[]{
\begin{tikzpicture}
\begin{feynman}
\vertex(a);
\vertex[below=0.5cm of a] (b);
\vertex[below=0.5cm of b] (c);
\vertex[below left=0.5cm and 1cm of a] (a3){\(\chi\)};
\vertex[above left=0.5cm and 1cm of a] (a1) {\(\chi\)};
\vertex[above right=0.5cm and 1cm of a] (a2){\(\chi\)};
\vertex[below right=0.5cm and 1cm of c] (c2){\(\chi\)};
\vertex[below left=0.5cm and 1cm of c] (c1){\(\chi\)};
\diagram*{
(a) -- [ line width=0.25mm,charged scalar, arrow size=0.7pt,style=gray!50,edge label={\(\rm\textcolor{black}{ \chi}\)}] (b),
(c) -- [ line width=0.25mm,charged scalar, arrow size=0.7pt,edge label={\(\rm \)},style=gray!50,edge label'={\(\rm\textcolor{black}{ \chi}\)}] (b),
(a) -- [ line width=0.25mm,charged scalar, arrow size=0.7pt,edge label={\(\rm \)},style=black] (c1),
(a1) -- [ line width=0.25mm,charged scalar, arrow size=0.7pt,edge label={\(\rm \)},style=black] (b),
(c) -- [ line width=0.25mm,charged scalar, arrow size=0.7pt,edge label={\(\rm \)},style=black] (a3),
(a) -- [ line width=0.25mm,charged scalar, arrow size=0.7pt,edge label={\(\rm \)},style=red] (a2),
(c) -- [ line width=0.25mm,charged scalar, arrow size=0.7pt,edge label={\(\rm \)},style=red] (c2)};
\end{feynman}
\end{tikzpicture}\label{fig:2i}}
\caption{Self-annihilation of SIMP via $\chi \chi^* \chi^* \rightarrow \chi \chi$ process.}
\label{fig:2}
\end{figure}
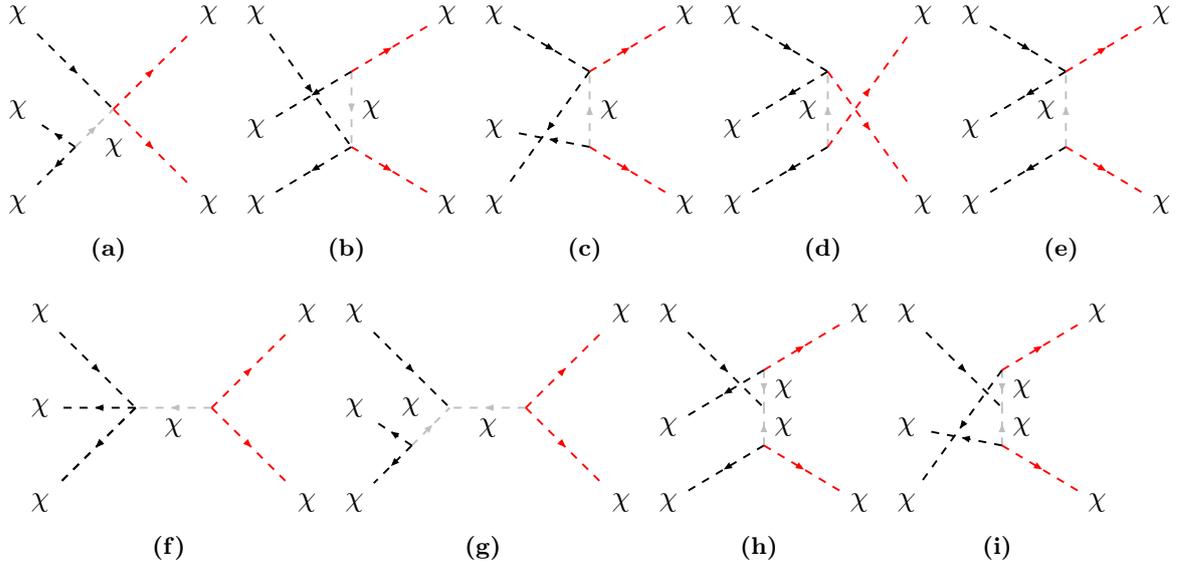

\begin{gather}
\overline{|\mathcal{M}_{\ref{fig:2}}|^2}=\dfrac{\mu_3^2}{64m_{\chi}^8}\left(117\mu_3^2-148\lambda_{\chi}m_{\chi}^2\right)^2\,.
\end{gather}
and
\begin{gather}
\langle\sigma v^2\rangle_{\chi\chi^*\chi^*\to\chi\chi}=\dfrac{1}{64\pi m_{\chi}^3}\left(\dfrac{K_1(m_{\chi}/T)}{K_2(m_{\chi}/T)}\right)^3\dfrac{\sqrt{5}}{6}\overline{|\mathcal{M}_{\chi\chi^*\chi^*\to\chi\chi}|^2}\,.
\end{gather}

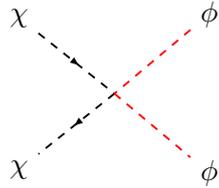
\begin{figure}[htb!]
\centering
\begin{tikzpicture}
\begin{feynman}
\vertex(a1);
\vertex[above left=0.75cm and 1cm of a1] (a2){\(\chi\)};
\vertex[below left=0.75cm and 1cm of a1] (a3){\(\chi\)};
\vertex[right=1cm of a1] (b1);
\vertex[above right=0.75cm and 1cm of a1] (b2){\(\phi\)};
\vertex[below right=0.75cm and 1cm of a1] (b3){\(\phi\)};
\diagram*{
(a2) -- [ line width=0.25mm,charged scalar, arrow size=0.7pt] (a1) -- [ line width=0.25mm,charged scalar, arrow size=0.7pt] (a3) ,
(b2) -- [ line width=0.25mm,scalar, style=red, edge label={\(\rm\color{black}{ }\)}] (a1) -- [ line width=0.25mm,scalar, style=red, arrow size=0.7pt] (b3)};
\end{feynman}
\end{tikzpicture}
\caption{Conversion of SIMP to pFIMP via $\chi \chi^* \rightarrow \phi \phi$ process.}
\label{fig:conversion}
\end{figure}

\section{Dark Matter self-interaction}
\label{app:C}
In single component complex scalar SIMP scenario, The effective DM self-scattering cross-section over mass is written as \cite{Choi:2015bya},
\begin{align}
\frac{\sigma_{\rm self}}{m_{\rm DM}}=\frac{1}{m_{\chi}}\left(\dfrac{\Omega^2_{\chi}}{\Omega_{\rm DM}^2}\sigma_{\chi\chi\to\chi\chi}+\dfrac{\Omega^2_{\chi^*}}{\Omega_{\rm DM}^2}\sigma_{\chi^{*}\chi^{*}\to\chi^{*}\chi^{*}}+\dfrac{\Omega_{\chi}}{\Omega_{\rm DM}}\dfrac{\Omega_{\chi^*}}{\Omega_{\rm DM}}\sigma_{\chi\chi^{*}\to\chi\chi^{*}}\right)\,.
\label{eq:simp_dm-self}
\end{align}
\begin{figure}[htb!]
\centering
\subfloat[]{\begin{tikzpicture}
\begin{feynman}
\vertex(a);
\vertex[above left =0.5cm and 0.75cm of a] (a1){\(\chi\)};
\vertex[below left =0.5cm and 0.75cm of a] (a2){\(\chi\)};
\vertex[above right =0.5cm and 0.75cm of a] (a3){\(\chi\)};
\vertex[below right =0.5cm and 0.75cm of a] (a4){\(\chi\)};
\diagram*{(a1) -- [ line width=0.25mm,charged scalar, arrow size=0.7pt,edge label={\(\rm \)},style=black] (a),(a2) -- [ line width=0.25mm,charged scalar, arrow size=0.7pt,edge label={\(\rm \)},style=black] (a),(a) -- [ line width=0.25mm,charged scalar, arrow size=0.7pt,style=red] (a3),(a)-- [ line width=0.25mm,charged scalar, arrow size=0.7pt,style=red] (a4)};
\end{feynman}
\end{tikzpicture}}\subfloat[]{
\begin{tikzpicture}
\begin{feynman}
\vertex(a);
\vertex[right=1cm of a] (b);
\vertex[above left =0.5cm and 0.75cm of a] (a1){\(\chi\)};
\vertex[below left =0.5cm and 0.75cm of a] (a2){\(\chi\)};
\vertex[above right =0.5cm and 0.75cm of b] (b1){\(\chi\)};
\vertex[below right =0.5cm and 0.75cm of b] (b2){\(\chi\)};
\diagram*{(a1) -- [ line width=0.25mm,charged scalar, arrow size=0.7pt,edge label={\(\rm \)},style=black] (a) -- [ line width=0.25mm,charged scalar, arrow size=0.7pt,edge label={\(\rm \)},style=black] (a2),(b) -- [ line width=0.25mm,charged scalar, arrow size=0.7pt,style=red] (b1),(b)-- [ line width=0.25mm,charged scalar, arrow size=0.7pt,style=red] (b2),(b) -- [ line width=0.25mm,charged scalar, arrow size=0.7pt,style=gray!50,edge label={\(\rm\textcolor{black}{ \chi}\)}] (a)};
\end{feynman}
\end{tikzpicture}}\subfloat[]{
\begin{tikzpicture}
\begin{feynman}
\vertex(a);
\vertex[below=1.5cm of a] (b);
\vertex[left=0.9cm of a] (a1){\(\chi\)};
\vertex[left=0.9cm of b] (b1){\(\chi\)};
\vertex[right=0.9cm of a] (a2){\(\chi\)};
\vertex[right=0.9cm of b] (b2){\(\chi\)};
\diagram*{(a1) -- [ line width=0.25mm,charged scalar, arrow size=0.7pt,edge label={\(\rm \)},style=black] (a) -- [ line width=0.25mm,charged scalar, arrow size=0.7pt,edge label={\(\rm \)},style=red] (a2),(b1) -- [ line width=0.25mm,charged scalar, arrow size=0.7pt,style=black] (b)-- [ line width=0.25mm,charged scalar, arrow size=0.7pt,style=red] (b2),(a) -- [ line width=0.25mm,scalar,style=gray!50,edge label={\(\rm\textcolor{black}{ h}\)}] (b) };
\end{feynman}
\end{tikzpicture}}\subfloat[]{
\begin{tikzpicture}
\begin{feynman}
\vertex(a);
\vertex[below=1.5cm of a] (b);
\vertex[left=0.9cm of a] (a1){\(\chi\)};
\vertex[left=0.9cm of b] (b1){\(\chi\)};
\vertex[right=1.5cm of a] (a2){\(\chi\)};
\vertex[right=1.5cm of b] (b2){\(\chi\)};
\diagram*{(a1) -- [ line width=0.25mm,charged scalar, arrow size=0.7pt,edge label={\(\rm \)},style=black] (a) -- [ line width=0.25mm,charged scalar, arrow size=0.7pt,edge label={\(\rm \)},style=red] (b2),(b1) -- [ line width=0.25mm,charged scalar, arrow size=0.7pt,style=black] (b)-- [ line width=0.25mm,charged scalar, arrow size=0.7pt,style=red] (a2),(a) -- [ line width=0.25mm,scalar,style=gray!50,edge label={\(\rm\textcolor{black}{ h}\)}] (b) };
\end{feynman}
\end{tikzpicture}}\subfloat[]{
\begin{tikzpicture}
\begin{feynman}
\vertex(a);
\vertex[above left =0.5cm and 0.75cm of a] (a1){\(\chi\)};
\vertex[below left =0.5cm and 0.75cm of a] (a2){\(\chi\)};
\vertex[above right =0.5cm and 0.75cm of a] (a3){\(\chi\)};
\vertex[below right =0.5cm and 0.75cm of a] (a4){\(\chi\)};
\diagram*{(a1) -- [ line width=0.25mm,charged scalar, arrow size=0.7pt,edge label={\(\rm \)},style=black] (a),(a) -- [ line width=0.25mm,charged scalar, arrow size=0.7pt,edge label={\(\rm \)},style=black] (a2),(a) -- [ line width=0.25mm,charged scalar, arrow size=0.7pt,style=red] (a3),(a4)-- [ line width=0.25mm,charged scalar, arrow size=0.7pt,style=red] (a)};
\end{feynman}
\end{tikzpicture}}

\subfloat[]{
\begin{tikzpicture}
\begin{feynman}
\vertex(a);
\vertex[right=1cm of a] (b);
\vertex[above left =0.5cm and 0.75cm of a] (a1){\(\chi\)};
\vertex[below left =0.5cm and 0.75cm of a] (a2){\(\chi\)};
\vertex[above right =0.5cm and 0.75cm of b] (b1){\(\chi\)};
\vertex[below right =0.5cm and 0.75cm of b] (b2){\(\chi\)};
\diagram*{(a1) -- [ line width=0.25mm,charged scalar, arrow size=0.7pt,edge label={\(\rm \)},style=black] (a) -- [ line width=0.25mm,charged scalar, arrow size=0.7pt,edge label={\(\rm \)},style=black] (a2),(b) -- [ line width=0.25mm,charged scalar, arrow size=0.7pt,style=red] (b1),(b2)-- [ line width=0.25mm,charged scalar, arrow size=0.7pt,style=red] (b),(b) -- [ line width=0.25mm, scalar, arrow size=0.7pt,style=gray!50,edge label={\(\rm\textcolor{black}{ h}\)}] (a)};
\end{feynman}
\end{tikzpicture}}
\subfloat[]{
\begin{tikzpicture}
\begin{feynman}
\vertex(a);
\vertex[below=1.5cm of a] (b);
\vertex[left=0.9cm of a] (a1){\(\chi\)};
\vertex[left=0.9cm of b] (b1){\(\chi\)};
\vertex[right=0.9cm of a] (a2){\(\chi\)};
\vertex[right=0.9cm of b] (b2){\(\chi\)};
\diagram*{(a1) -- [ line width=0.25mm,charged scalar, arrow size=0.7pt,edge label={\(\rm \)},style=black] (a) -- [ line width=0.25mm,charged scalar, arrow size=0.7pt,edge label={\(\rm \)},style=red] (a2),(b2) -- [ line width=0.25mm,charged scalar, arrow size=0.7pt,style=red] (b)-- [ line width=0.25mm,charged scalar, arrow size=0.7pt,style=black] (b1),(a) -- [ line width=0.25mm, scalar,style=gray!50,edge label={\(\rm\textcolor{black}{ h}\)}] (b) };
\end{feynman}
\end{tikzpicture}}\subfloat[]{
\begin{tikzpicture}
\begin{feynman}
\vertex(a);
\vertex[below=1.5cm of a] (b);
\vertex[left=0.9cm of a] (a1){\(\chi\)};
\vertex[left=0.9cm of b] (b1){\(\chi\)};
\vertex[right=0.9cm of a] (a2){\(\chi\)};
\vertex[right=0.9cm of b] (b2){\(\chi\)};
\diagram*{(a1) -- [ line width=0.25mm,charged scalar, arrow size=0.7pt,edge label={\(\rm \)},style=black] (a),(a2) -- [ line width=0.25mm,charged scalar, arrow size=0.7pt,edge label={\(\rm \)},style=red] (a),(b) -- [ line width=0.25mm,charged scalar, arrow size=0.7pt,style=black] (b1),(b)-- [ line width=0.25mm,charged scalar, arrow size=0.7pt,style=red] (b2),(a) -- [ line width=0.25mm,charged scalar, arrow size=0.7pt,style=gray!50,edge label={\(\rm\textcolor{black}{ \chi}\)}] (b) };
\end{feynman}
\end{tikzpicture}}\subfloat[]{
\begin{tikzpicture}
\begin{feynman}
\vertex(a);
\vertex[below=1.5cm of a] (b);
\vertex[left=0.9cm of a] (a1){\(\phi\)};
\vertex[left=0.9cm of b] (b1){\(\chi\)};
\vertex[right=0.9cm of a] (a2){\(\phi\)};
\vertex[right=0.9cm of b] (b2){\(\chi\)};
\diagram*{(a1) -- [ line width=0.25mm, scalar, arrow size=0.7pt,edge label={\(\rm \)},style=black] (a) -- [ line width=0.25mm, scalar, arrow size=0.7pt,edge label={\(\rm \)},style=red] (a2),(b) -- [ line width=0.25mm,charged scalar, arrow size=0.7pt,style=black] (b1),(b2)-- [ line width=0.25mm,charged scalar, arrow size=0.7pt,style=red] (b),(a) -- [ line width=0.25mm,scalar,style=gray!50,edge label={\(\rm\textcolor{black}{ h}\)}] (b) };
\end{feynman}
\end{tikzpicture}}\subfloat[]{
\begin{tikzpicture}
\begin{feynman}
\vertex(a);
\vertex[below=1.5cm of a] (b);
\vertex[left=0.9cm of a] (a1){\(\phi\)};
\vertex[left=0.9cm of b] (b1){\(\chi\)};
\vertex[right=0.9cm of a] (a2){\(\phi\)};
\vertex[right=0.9cm of b] (b2){\(\chi\)};
\diagram*{(a1) -- [ line width=0.25mm, scalar, arrow size=0.7pt,edge label={\(\rm \)},style=black] (a) -- [ line width=0.25mm, scalar, arrow size=0.7pt,edge label={\(\rm \)},style=red] (a2),(b1) -- [ line width=0.25mm,charged scalar, arrow size=0.7pt,style=black] (b)-- [ line width=0.25mm,charged scalar, arrow size=0.7pt,style=red] (b2),(a) -- [ line width=0.25mm,scalar,style=gray!50,edge label={\(\rm\textcolor{black}{ h}\)}] (b) };
\end{feynman}
\end{tikzpicture}}

\subfloat[]{
\begin{tikzpicture}
\begin{feynman}
\vertex(a);
\vertex[above left =0.5cm and 0.75cm of a] (a1){\(\phi\)};
\vertex[below left =0.5cm and 0.75cm of a] (a2){\(\chi\)};
\vertex[above right =0.5cm and 0.75cm of a] (a3){\(\chi\)};
\vertex[below right =0.5cm and 0.75cm of a] (a4){\(\phi\)};
\diagram*{(a1) -- [ line width=0.25mm, scalar, arrow size=0.7pt,edge label={\(\rm \)},style=black] (a) -- [ line width=0.25mm,charged scalar, arrow size=0.7pt,edge label={\(\rm \)},style=black] (a2),(a3) -- [ line width=0.25mm,charged scalar, arrow size=0.7pt,style=red] (a),(a)-- [ line width=0.25mm, scalar, arrow size=0.7pt,style=red] (a4)};
\end{feynman}
\end{tikzpicture}}\subfloat[]{
\begin{tikzpicture}
\begin{feynman}
\vertex(a);
\vertex[above left =0.5cm and 0.75cm of a] (a1){\(\phi\)};
\vertex[below left =0.5cm and 0.75cm of a] (a2){\(\chi\)};
\vertex[above right =0.5cm and 0.75cm of a] (a3){\(\chi\)};
\vertex[below right =0.5cm and 0.75cm of a] (a4){\(\phi\)};
\diagram*{(a1) -- [ line width=0.25mm, scalar, arrow size=0.7pt,edge label={\(\rm \)},style=black] (a),(a2) -- [ line width=0.25mm,charged scalar, arrow size=0.7pt,edge label={\(\rm \)},style=black] (a) -- [ line width=0.25mm,charged scalar, arrow size=0.7pt,style=red] (a3),(a)-- [ line width=0.25mm, scalar, arrow size=0.7pt,style=red] (a4)};
\end{feynman}
\end{tikzpicture}}
\subfloat[]{
\begin{tikzpicture}
\begin{feynman}
\vertex(a);
\vertex[above left =0.5cm and 0.75cm of a] (a1){\(\phi\)};
\vertex[below left =0.5cm and 0.75cm of a] (a2){\(\phi\)};
\vertex[above right =0.5cm and 0.75cm of a] (a3){\(\phi\)};
\vertex[below right =0.5cm and 0.75cm of a] (a4){\(\phi\)};
\diagram*{(a1) -- [ line width=0.25mm, scalar, arrow size=0.7pt,edge label={\(\rm \)},style=black] (a),(a) -- [ line width=0.25mm, scalar, arrow size=0.7pt,edge label={\(\rm \)},style=black] (a2),(a) -- [ line width=0.25mm, scalar, arrow size=0.7pt,style=red] (a3),(a4)-- [ line width=0.25mm, scalar, arrow size=0.7pt,style=red] (a)};
\end{feynman}
\end{tikzpicture}}\subfloat[]{
\begin{tikzpicture}
\begin{feynman}
\vertex(a);
\vertex[right=1cm of a] (b);
\vertex[above left =0.5cm and 0.75cm of a] (a1){\(\phi\)};
\vertex[below left =0.5cm and 0.75cm of a] (a2){\(\phi\)};
\vertex[above right =0.5cm and 0.75cm of b] (b1){\(\phi\)};
\vertex[below right =0.5cm and 0.75cm of b] (b2){\(\phi\)};
\diagram*{(a1) -- [ line width=0.25mm, scalar, arrow size=0.7pt,edge label={\(\rm \)},style=black] (a) -- [ line width=0.25mm, scalar, arrow size=0.7pt,edge label={\(\rm \)},style=black] (a2),(b) -- [ line width=0.25mm, scalar, arrow size=0.7pt,style=red] (b1),(b2)-- [ line width=0.25mm, scalar, arrow size=0.7pt,style=red] (b),(b) -- [ line width=0.25mm, scalar, arrow size=0.7pt,style=gray!50,edge label={\(\rm\textcolor{black}{ h}\)}] (a)};
\end{feynman}
\end{tikzpicture}}\subfloat[]{
\begin{tikzpicture}
\begin{feynman}
\vertex(a);
\vertex[below=1.5cm of a] (b);
\vertex[left=0.9cm of a] (a1){\(\phi\)};
\vertex[left=0.9cm of b] (b1){\(\phi\)};
\vertex[right=0.9cm of a] (a2){\(\phi\)};
\vertex[right=0.9cm of b] (b2){\(\phi\)};
\diagram*{(a1) -- [ line width=0.25mm, scalar, arrow size=0.7pt,edge label={\(\rm \)},style=black] (a) -- [ line width=0.25mm, scalar, arrow size=0.7pt,edge label={\(\rm \)},style=red] (a2),(b2) -- [ line width=0.25mm, scalar, arrow size=0.7pt,style=red] (b)-- [ line width=0.25mm, scalar, arrow size=0.7pt,style=black] (b1),(a) -- [ line width=0.25mm, scalar,style=gray!50,edge label={\(\rm\textcolor{black}{ h}\)}] (b) };
\end{feynman}
\end{tikzpicture}}
\caption{Dark matter self scattering}\label{self-scattering}
\end{figure}
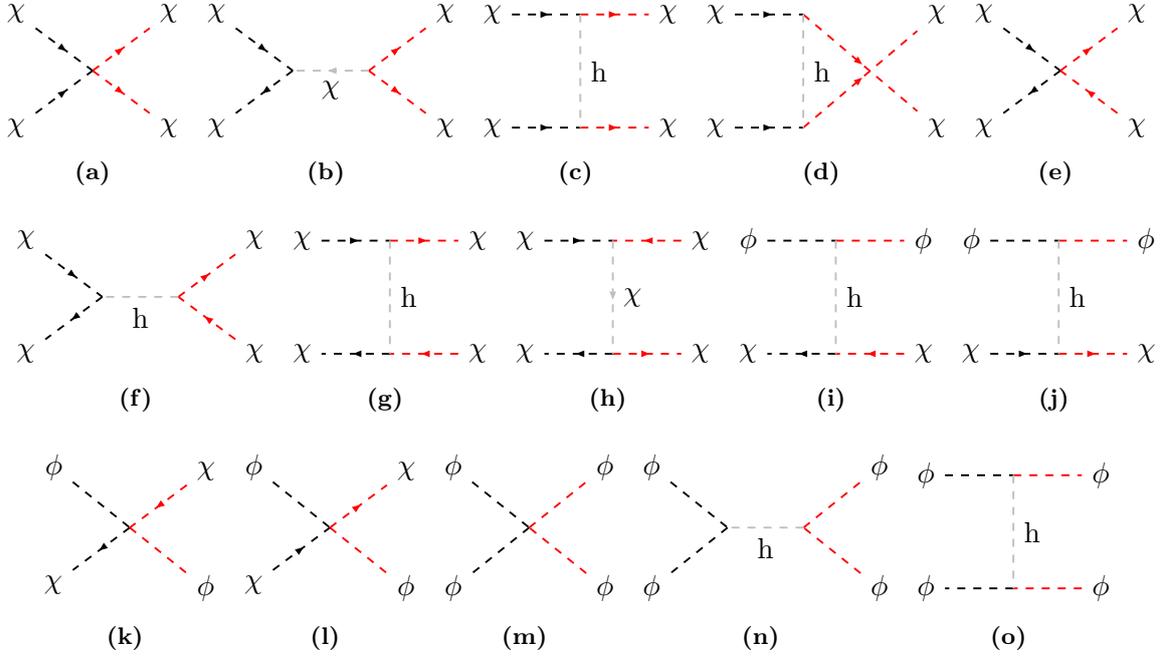
In the multicomponent scenario, we take DM mass as effective $\rm m_{\rm DM}$ weighted by effective relic contribution for a particular DM and the analytic expression could be written as \cite{Choi:2021yps},
\bea\begin{split}
\frac{\sigma_{\rm self}}{m_{\rm DM}}=&\left(\frac{\Omega_{\chi}}{\Omega_{\rm DM}}\right)^2\frac{1}{m_{\chi}}\left(\sigma_{\chi\chi\to\chi\chi}+\sigma_{\chi^{*}\chi^{*}\to\chi^{*}\chi^{*}}+\sigma_{\chi\chi^{*}\to\chi\chi^{*}}+\sigma_{\chi\chi^{*}\to\phi\phi}\right)\\&+\left(\frac{\Omega_{\phi}}{\Omega_{\rm DM}}\right)^2\frac{1}{m_{\phi}}\sigma_{\phi\phi\to\phi\phi}+\frac{\Omega_{\chi}}{\Omega_{\rm DM}}\frac{\Omega_{\phi}}{\Omega_{\rm DM}}\frac{2}{m_{\chi}+m_{\phi}}\left(\sigma_{\phi\chi\to \phi\chi}+\sigma_{\phi\chi^{*}\to \phi\chi^{*}}\right)\,.
\label{eq:dm-self}
\end{split}\eea
where,
\begin{align}
\sigma_{\chi\chi\to\chi\chi}&=\dfrac{1}{16\pi (m_{\chi}+m_{\chi})^2}\left(|\mathcal{M}_{\chi\chi\to\chi\chi}|^2+|\mathcal{M}_{\chi^*\chi^*\to\chi^*\chi^*}|^2\right)\,,\\
\sigma_{\chi\chi^*\to\chi\chi^*}&=\dfrac{1}{16\pi (m_{\chi}+m_{\chi^*})^2}|\mathcal{M}_{\chi\chi^*\to\chi\chi^*}|^2\,,\\\
\sigma_{\chi\chi^*\to\phi\phi}&=\dfrac{1}{16\pi (m_{\chi}+m_{\chi^*})^2}|\mathcal{M}_{\chi\chi^*\to\phi\phi}|^2\,,\\
\sigma_{\phi\phi\to\phi\phi}&=\dfrac{1}{16\pi (m_{\phi}+m_{\phi})^2}|\mathcal{M}_{\phi\phi\to\phi\phi}|^2\,,\\
\sigma_{\phi\chi\to\phi\chi}&=\dfrac{1}{16\pi (m_{\phi}+m_{\chi})^2}\left(|\mathcal{M}_{\phi\chi\to\phi\chi}|^2+|\mathcal{M}_{\phi\chi^*\to\phi\chi^*}|^2\right)\,,
\end{align} 
with,
\begin{align}
|\mathcal{M}_{\chi\chi\to\chi\chi}|^2&=\frac{\left(m_h^2(3 \mu_3^2+4\lambda_{\chi}m_{\chi}^2)-2\lambda_{\chi H}^2v^2m_{\chi}^2\right)^2}{2 m_{\chi}^4m_h^4}\,,\\
|\mathcal{M}_{\chi\chi^*\to\chi\chi^*}|^2&=\frac{\left(m_h^2(4m_{\chi}^2-m_h^2)(9 \mu_3^2-4\lambda_{\chi}m_{\chi}^2)-2\lambda_{\chi H}^2v^2m_{\chi}^2(m_h^2-2m_{\chi}^2)\right)^2}{m_{\chi}^4m_h^4(m_h^2-4m_{\chi}^2)^2}\,,\\
|\mathcal{M}_{\chi\chi^*\to\phi\phi}|^2&=\frac{\left(4\lambda_{\chi\phi}m_{\chi}^2-\lambda_{\chi\phi}m_h^2+\lambda_{\chi H}\lambda_{\phi H}v^2\right)^2}{2(m_h^2-4m_{\chi}^2)^2}\,,\\
|\mathcal{M}_{\chi\phi\to\chi\phi}|^2&=\frac{\left(\lambda_{\chi\phi}\left((m_{\chi}-m_{\phi})^2-m_h^2\right)-\lambda_{\chi H}\lambda_{\phi H}v^2\right)^2}{\left((m_{\chi}-m_{\phi})^2-m_h^2\right)^2}\,,\\
|\mathcal{M}_{\phi\phi\to\phi\phi}|^2&=\frac{\left(\lambda_{\phi}(m_h^4-4m_h^2m_{\phi}^2)+\lambda_{\phi H}^2(8m_{\phi}^2-3m_{h}^2)v^2\right)^2}{2m_h^4(m_h^2-4m_{\phi}^2)^2}\,.
\end{align}
and $\Omega_{\chi}=\Omega_s/2$, $\Omega_{\rm DM}=\Omega_s+\Omega_{\phi}.$
\section{DM Kinetic equilibriation}
\label{app:D}
\begin{figure}[htb!]
\centering
\begin{tikzpicture}
	\begin{feynman}
	\vertex (a);
	\vertex[below=1.5cm of a] (b);
	\vertex[left=1cm of a] (a1){\(\chi\)};
	\vertex[left=1cm of b] (b1){\(f\)};
	\vertex[right=1cm of a] (a2){\(\chi\)};
	\vertex[right=1cm of b] (b2){\(f\)};
	\diagram*{(a1) -- [ line width=0.25mm,charged scalar, arrow size=0.7pt,edge label={\(\rm \)},style=black] (a) -- [ line width=0.25mm,charged scalar, arrow size=0.7pt,edge label={\(\rm \)},style=red] (a2),(b1) -- [ line width=0.25mm,fermion, arrow size=0.7pt,style=black] (b)-- [ line width=0.25mm,fermion, arrow size=0.7pt,style=red] (b2),(a) -- [ line width=0.25mm,scalar,style=gray!50,edge label={\(\rm\textcolor{black}{ h}\)}] (b) };
	\end{feynman}
	\end{tikzpicture}
	\caption{The Feynman diagram represent the DM $(\chi)$ and SM fermions $(f)$ scattering.}
\label{fig:26a}
\end{figure}
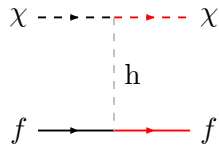

We already discussed that the kinetic equilibrium of SIMP with SM is achieved after considering an appropriate choice of portal coupling. Only relativistic fermions are available during SIMP freezes-out, and contributing to kinetic decoupling condition is written regarding DM-SM scattering, fig\,.~\ref{fig:26a}, rate as,
\bea
2\sum_{f}\Gamma_{\chi f\to \chi\to f}=2\sum_{f}\langle \sigma v\rangle_{\chi f\to \chi f}n_f^{\rm eq}(T)\gtrsim \mathcal{H}(T)\,.
\label{eq:kin-decup}
\eea
where,
\begin{align}
\langle \sigma v\rangle_{\chi f\to\chi f}=\frac{\int_{(m_{\chi}+m_{f})^2}^{\infty}\sigma_{\chi f\to\chi f}K_1\left(\frac{\sqrt{s}}{T}\right)\sqrt{s-(m_{\chi}-m_f)^2}\left(s-(m_{\chi}+m_{f})^2\right)ds}{8Tm_{\chi}^2m_{f}^2K_2\left(\frac{m_{\chi}}{T}\right)K_2\left(\frac{m_{f}}{T}\right)}\,.
\end{align}
and two factors come from SM fermion and anti-fermion contribution.
\begin{figure}[htb!]
\centering
\includegraphics[width=0.475\linewidth]{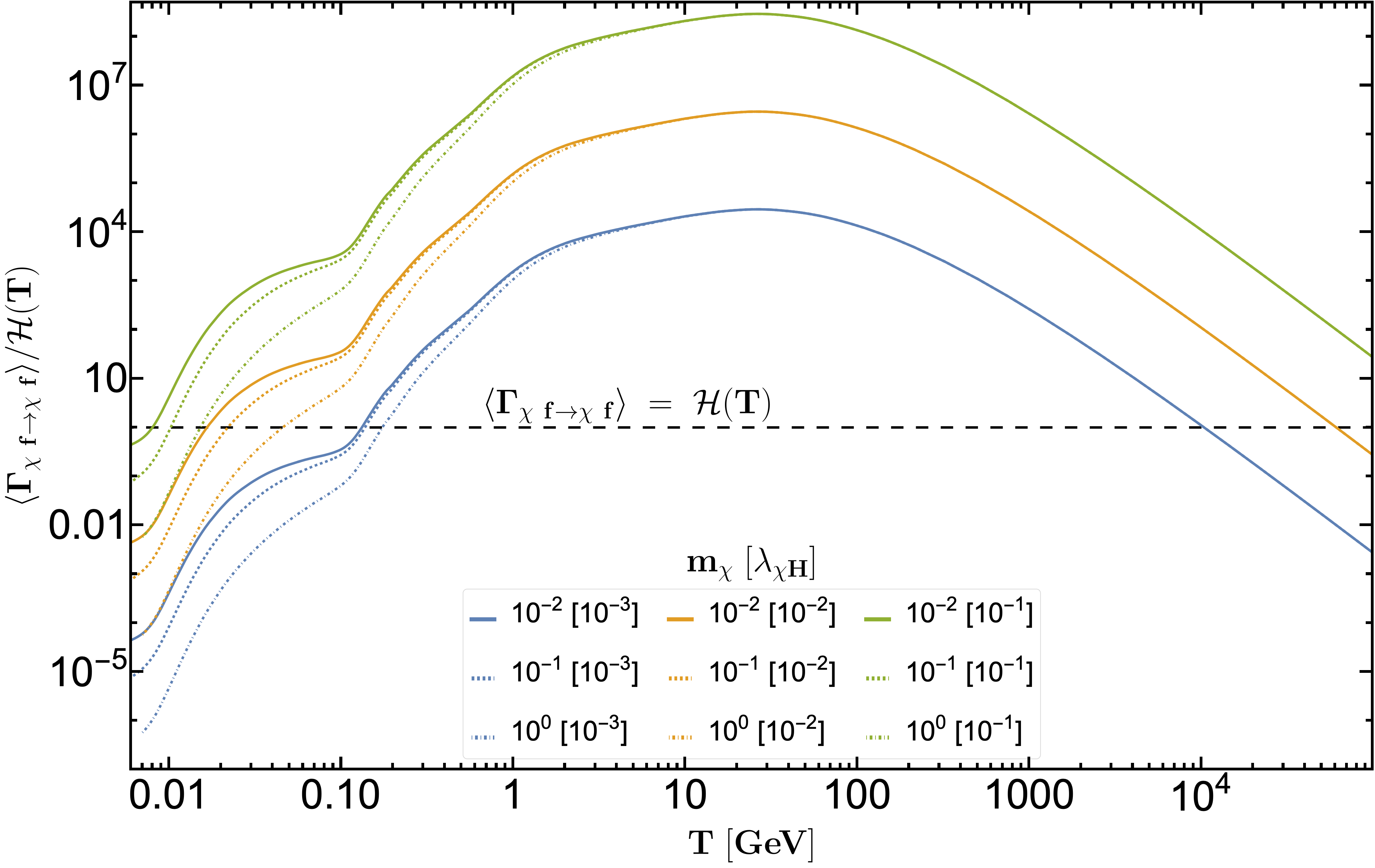}
\caption{The evaluation of DM $(\chi)$-fermion elastic scattering rate with temperature. The different color lines represent the different $m_{\chi}$ masses in the GeV unit, and coupling is mentioned in the figure inset.}
\label{fig:kint-decup}
\end{figure}
Fig\,.~\ref{fig:kint-decup} represents the evolution of DM-SM elastic scattering rate with the SM bath temperature $(T)$. Our work focuses on sub-MeV DMs and suitable choice of $\lambda_{\chi H}\sim 0.1$, without affecting DM relic, to achieve kinetic equilibrium between DM and SM bath. 

The kinetic equilibriation, with thermal bath, of $\chi$ ensures the pFIMP ($\phi$) kinetic equilibriation also because of the presence of strong self-interaction among themselves. For this reason, there is no necessity to do similar kinds of analysis for $\phi$.

\newpage
\bibliographystyle{JHEP}
\bibliography{main}

\end{document}